\documentclass[usegraphicx,usenatbib]{mn2e}

\newcommand{\Halpha}{\ensuremath{H_{\alpha}}}

\newcommand{\Hzero}{\ensuremath{H_{\mathrm{0}}}}

\newcommand{\LB}{\ensuremath{L_{\mathrm{B}}}}

\newcommand{\LFIR}{\ensuremath{L_{\mathrm{FIR}}}}

\newcommand{\LK}{\ensuremath{L_{\mathrm{K}}}}
\newcommand{\Lsol}{\ensuremath{\mathrm{L}_{\odot}}}
\newcommand{\LULX}{\ensuremath{L_{\mathrm{ULX}}}}
\newcommand{\LX}{\ensuremath{L_{\mathrm{X}}}}

\newcommand{\Msol}{\ensuremath{\mathrm{M}_{\odot}}}

\newcommand{\NH}{\ensuremath{N_{\mathrm{H}}}}

\newcommand{\NULX}{\ensuremath{N(\mathrm{ULX})}}

\newcommand{\rSN}{\ensuremath{r_{\mathrm{SN}}}}

\newcommand{\Z}{\ensuremath{Z}}
\newcommand{\Zsol}{\ensuremath{\mathrm{Z}_{\odot}}}

\newcommand{\arcm}{\ensuremath{^\prime}}
\newcommand{\arcs}{\ensuremath{^{\prime\prime}}}

\newcommand{\Mpc}{\ensuremath{\mbox{Mpc}}}

\newcommand{\nm}{\ensuremath{\mbox{\nm}}}
\newcommand{\pc}{\ensuremath{\mbox{pc}}}

\newcommand{\CHANDRA}{\emph{Chandra}}

\newcommand{\ROSAT}{\emph{ROSAT}}

\title{{\em Chandra} Observations of the Interacting Galaxies NGC 3395/3396 (Arp 270)}

\author[Brassington et al.] 
  {Nicola J. Brassington,$^1$\thanks{E-mail: njb@star.sr.bham.ac.uk} Andrew M. Read$^2$ and Trevor J. Ponman$^1$ \\
  $^1$School of Physics and Astronomy, The University of Birmingham,
  Edgbaston, Birmingham B15 2TT, UK\\
  $^2$Department of Physics and Astronomy, University of Leicester,University Road, Leicester LE1 7RH, UK}

\date{Accepted 2005 ??. Received 2005 ??; in original form 2004 ??}

\begin{document}

\maketitle

\begin{abstract}

In this paper we present the results of a 20 ks high resolution
Chandra X-ray observation of the peculiar galaxy pair NGC 3395/3396, a
system at a very early stage of merging, and less evolved than
the famous Antennae and Mice merging systems. Previously unpublished
ROSAT HRI data are also presented. The point source
population and the hot diffuse gas in this system are investigated,
and compared with other merging galaxy pairs.

16 X-ray point sources are detected in Arp 270, 7 of which are
classified as ULXs (\LX\ $\ge 10^{39}$erg s$^{-1}$). From
spectral fits and the age of the system it seems likely that these are
predominantly high mass X-ray binaries. The diffuse gas emits at a global
temperature of $\sim$0.5 keV, consistent with temperatures observed in
other interacting systems, and we see no evidence of the starburst-driven hot gaseous outflows
seen in more evolved systems such as The Mice and The Antennae. It
is likely that these features are absent from Arp 270 as the gas has had
insufficient time to break out of the galaxy disks. 32\% of the luminosity
of Arp 270 arises from the diffuse gas in the system, this is low when
compared to later stage merging systems and gives further credence that
this is an early stage merger.

Comparing the ULX population of Arp 270 to other merging systems, we
derive a relationship between the star formation rate of the system,
indicated by \LFIR, and the number (\NULX) and luminosity (\LULX) of its
ULX population. We find  \NULX$\propto\LFIR^{0.18}$ and
\LULX$\propto\LFIR^{0.54}$. These relationships, coupled with the relation
of the point source X-ray luminosity ($L_\mathrm{XP}$) to \LK\ and
$L_\mathrm{FIR+UV}$ \citep{Colbert_03}, indicate that the ULX sources in
an interacting system have contributions from both the old and young
stellar populations.

\end{abstract}

\begin{keywords} galaxies: evolution - galaxies: individual: NGC 3395 - galaxies: individual: NGC 3396 - galaxies: interactions - X-rays: galaxies

\end{keywords}

\section{Introduction}

It is widely believed some of the most dominant mechanisms in galaxy
evolution are galaxy collisions and mergers, and that there are very
few galaxies that have not been shaped or created through an
interaction or merger with another galaxy. As a consequence it could
be stated that the formation of many elliptical galaxies is the
product of the merger of two spiral galaxies. This was first proposed
by Toomre in 1977, illustrated by the `Toomre' sequence
\citep{Toomre_77}.  This shows 11 examples from the New General
Catalogue (NGC) of Nebulae and Clusters of Stars, from approaching
disk systems to near-elliptical remnants.  The systems within this sequence have been well studied over a range of wavelengths, including X-ray. Initial observations in X-ray were carried out with the {\em Einstein} Observatory, giving limited spatial resolution, these were later followed by observations with improved sensitivity from \ROSAT. With the advent of the
X-ray satellite {\em Chandra} \citep{Weisskopf_00}, these systems and other interacting galaxies
have been studied in much greater detail. 

With greater resolution the
X-ray properties of local galaxies can now be determined, and the
nature of the point sources in these systems established. Recent studies of starburst and merging galaxies have
included M82 \citep{Kaaret_01}, NGC 253 \citep{Strickland_00a}, NGC
3256 \citep{Lira_02}, NGC 4485 /4490 \citep{Roberts_02}, The Mice
\citep{Read_03}, and The Antennae \citep{Fabbiano_01}. In these papers
each system is found to have a population of luminous point sources
and the components of sub-structure in the diffuse emission have been
resolved. Most of these papers deal with the spectral analysis of the
point source population and the diffuse gas, allowing a comparison of
the system properties with the evolutionary stage it is in.

This paper reports a study of the galaxy pair NGC 3395/3396, which
appear in Arp's Atlas of Peculiar Galaxies \citep{Arp_66} as Arp 270
(Also VV 246; \citet{VV_59}). At a distance of 28 Mpc (assuming \Hzero\ = 75\,km s$^{-1}$ Mpc$^{-1}$, and accounting for Virgocentric in-fall) the galaxies have a separation of just 12kpc\ and, from a multicolour broad band photometric study, it has been shown that they are of
comparable mass \citep{Hern_01}, 
NGC 3395, to the
west, is classified as a Sc galaxy, and NGC 3396, to the east, as an
Irr.  From N-body simulations \citep{Clemens_99} it has been shown that Arp 270 is within 5$\times$10$^7$ years of its second perigalactic passage, the first occurring approximately 5$\times$10$^8$ years ago. Further evidence indicating that the two galaxies have interacted previously is the HI tail extending from the south-east of the galaxy \citep{Clemens_99}, although no optical tidal tails have been observed. This places the system at an earlier stage of evolution than the 11 examples given by \citet{Toomre_77}.

In this paper we present previous X-ray observations of Arp 270,
including previously unpublished results from the \ROSAT\ HRI
observation which are described in \textsection \ref{sec:pobs}. The
{\em Chandra} observations and data analysis techniques are detailed
in \textsection \ref{sec_data_red}. A full discussion of these results
is presented in \textsection \ref{sec_disc}. along with
multi-wavelength comparisons. Comparisons of the X-ray properties of
Arp 270 with other prominent starburst and merging systems are drawn
and discussed in \textsection \ref{sec_disc} and \textsection
\ref{conclusions}. Conclusions and summary are given in \textsection
\ref{conclusions}.


\section{Previous X-ray Observations}
\label{sec:pobs}


Arp~270 was initially observed in X-rays with the {\em Einstein} IPC
\citep{Fabbiano_92}. The image showed very little structure, however,
and what there was appeared to be centred on NGC~3395, with
suggestions of extensions to the northwest and southwest of the
system.


The {\em ROSAT} PSPC observations, described in \citet{Read_98}
(hereafter RP98), showed much more structure. Emission was seen from
both galaxies, and this appeared to lie within more extended,
diffuse gas. A point source was detected at the nucleus of
NGC~3395, while the central point source in NGC~3396 was seen, not at
the nucleus, but at a position closer to the companion
galaxy. Tentative evidence was also seen for complex emission in
the overlapping regions of the two disks, where there also appeared to
be quite a good correlation between the X-ray emission and the radio
emission of \citet{Huang_94}, with both an X-ray and a radio bridge
visible between the two galaxies. Not very much could be said as
regards the spectral properties of the emission, except that the
sources appeared to be best fitted with plasma models, suggestive of
the emission from both regions being caused primarily by hot gas,
rather than evolved stellar components. The remaining unresolved
emission appeared to be contaminated with point source
emission. However, a two-component fit did provide some evidence for
the existence of low-temperature gas.


Arp~270 was also observed with the {\em ROSAT} HRI. These previously
unpublished results are presented here, as they provide a useful
introduction to the X-ray structure of the system, and bridge the gap
between the (relatively) low-resolution of the ROSAT PSPC and the
higher resolution of Chandra.

The {\em ROSAT} HRI data (ID 600771), covering 24.9\,ks and taken in
May of 1995, were initially screened for good time intervals longer
than 10\,s. No further low background selection was made, as we were
mainly interested in point sources (which the HRI is far better suited
to), and for this purpose the data were still photon-limited. Source
detection and position determination was then performed over the full
field of view with the EXSAS local detect, map detect and maximum
likelihood algorithms \citep{Zimmermann_94}, using images of pixel
size 5\arcs.

\begin{figure}
  \includegraphics[height=\linewidth]{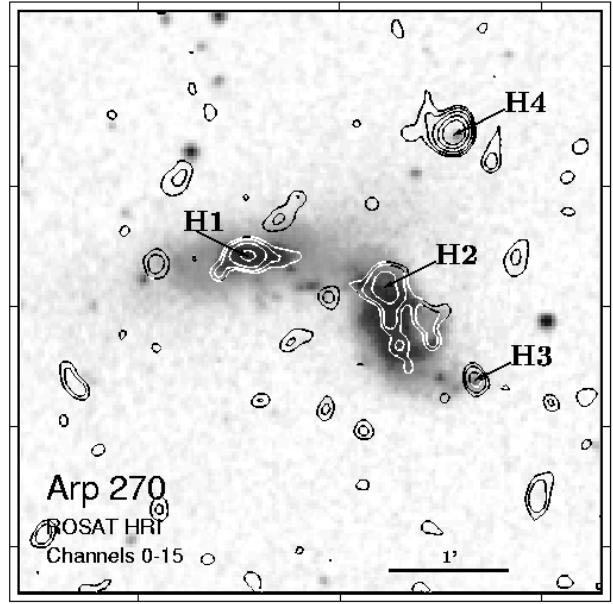}
  \caption {Contours of ROSAT HRI emission over the full channel range superimposed on an optical Digital Sky Survey (DSS) image of Arp~270. NGC 3395 lies to the west in the image, and NGC 3396 to the east. Labels indicate sources detected in the 0.1$-$2.0 keV band.  }
  \label{fig:hri}
\end{figure}

Figure \ref{fig:hri} shows smoothed contours of the {\em ROSAT} HRI
emission from Arp~270 over the full HRI channel range, superimposed
on an optical Digital Sky Survey (DSS) image. Sources were formally
detected at the positions of the two galaxies, embedded in regions of
spatially complex emission (most especially within NGC~3395). Two
other interesting sources were detected; a dim source at the
southwestern tip of NGC~3395, and a bright source to the northwest of
the galaxies.

The {\em ROSAT} HRI properties of these sources are summarised in
Table \ref{tab:hri} as follows: Source number column (1; prefixed by
H, for HRI), corrected right ascension and declination columns (2) and
(3), error on the source position column (4), and likelihood of
existence column (5). Probabilities P are related to maximum
likelihood values L by the relation $P=1-e^{-L}$, thus a likelihood L
of 10 corresponds to a Gaussian significance of 4.0$\sigma$
\citep{Zimmermann_94}. Also given are the net broad band counts and
error column (6), and count rates and errors after applying deadtime
and vignetting corrections column (7). The likelihood of the source
being extended and the associated extent are given in columns (8) and
(9). Note here that H1 (associated with NGC~3396) and especially H2
(NGC~3395) appear extended, whereas H3, to the SW of NGC~3395, and the
bright H4 to the NW of the system appear point-like.

\begin{table*}
\caption[]{
X-ray properties of ROSAT HRI-detected point sources
within and close to the optical confines of Arp~270
\label{tab:hri}}
\begin{tabular}{lcccccccc}
\noalign{\smallskip}
\hline
Src. & RA & Dec           & Pos.err.  & Lik. & Count(err)   & Ct.rate
& Lik.(ext) & extent \\
 &\multicolumn{2}{c}{(2000.0)} &(arcsec)&    &               &(ks$^{-1}$)
&           & (arcsec) \\ \hline

H1 & 10 49 55.65 & +32 59 25.8 & 1.6 &  22.0 & 29.47$\pm$6.5
&1.20$\pm$0.26 & 3.8 &  3.3 \\
H2 & 10 49 50.16 & +32 59 09.6 & 2.0 &  10.9 & 32.08$\pm$7.2
&1.30$\pm$0.29 & 9.0 &  4.9 \\
H3 & 10 49 46.64 & +32 58 23.4 & 1.3 &  13.6 & 13.01$\pm$4.2
&0.53$\pm$0.17 & 0.0 &   -  \\
H4 & 10 49 47.50 & +33 00 27.0 & 0.6 & 115.9 & 62.62$\pm$8.4
&2.55$\pm$0.34 & 0.7 &   -  \\
\noalign{\smallskip}
\hline
\end{tabular}
\end{table*}


\section{{\em Chandra} Observations and Data Analysis}
\label{sec_data_red}

Arp 270 was observed using the ACIS-S camera on board the {\em
Chandra} X-ray Observatory in two separate pointings on 28th April
2001 and 23rd October 2001 with a total observation time of
22.5\,ks. The initial data processing to correct for the motion of the
spacecraft and apply instrument calibration was carried out with the
Standard Data Processing (SDP) at the {\em Chandra} X-ray Center
(CXC). The data products were then analysed using the CXC CIAO
software suite (v2.3)\footnote{http://asc.harvard.edu/ciao} and
HEASOFT (v5.2). The data were reprocessed, screened for bad pixels, and time filtered to remove periods of high background (when
counts deviated by more than 5\( \sigma \) above the mean). This
resulted in a corrected exposure time of 20\,ks, 6.2\,ks in the first
pointing and 13.8\,ks in the second.

\subsection{Overall X-ray Structure}
\label{sec_struc}

\begin{figure*}
  \includegraphics[width=16cm]{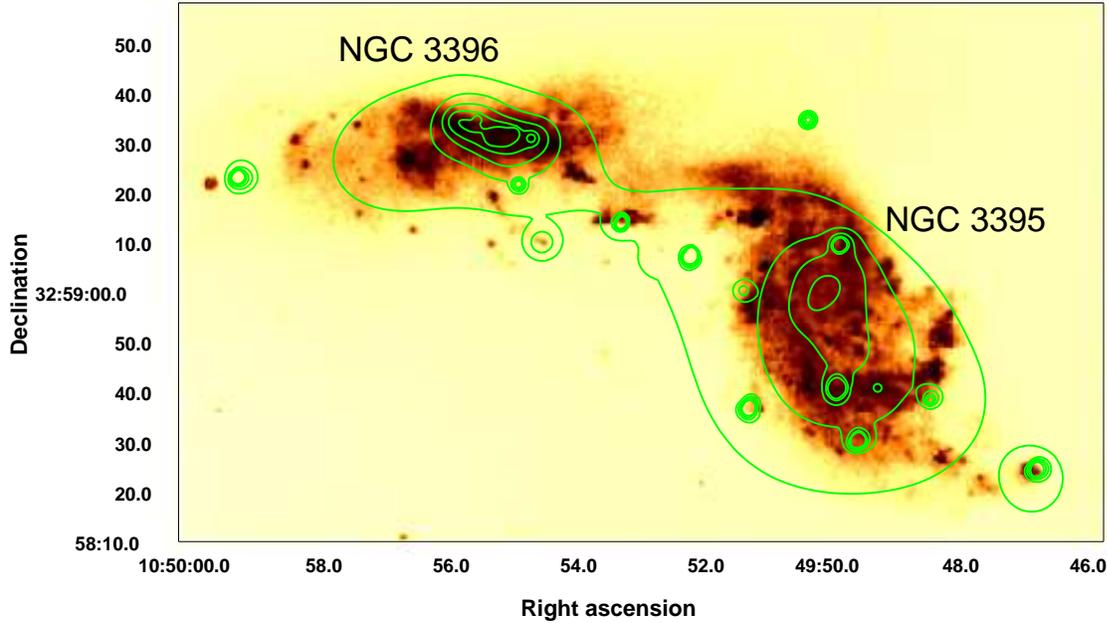}
  \hspace{0.1cm}
  \caption{Contours of adaptively smoothed 0.3$-$8.0\,keV X-ray data from {\em Chandra} ASIC-S overlaid on an optical image from the Palomar 5m telescope. Contours increase by a factor of $\sqrt{2}$. The nuclei of the galaxies are separated by 90 arcseconds. }
  \label{fig:optic_con}
\end{figure*}

A 0.3$-$8\,keV (from here on referred to as `full band') {\em Chandra}
image was created from the cleaned events file and adaptively smoothed
using the CIAO task {\em csmooth} which uses a smoothing kernel to
preserve an approximately constant signal to noise ratio across the
image, which was constrained to be between 2.4 and 4. Figure
\ref{fig:optic_con} shows an optical image of the interacting galaxies
from the Palomar 5m telescope with the full band X-ray contours
overlaid. From this image it can be seen that the galaxies are
resolved into two components. NGC 3395 shows diffuse gas throughout
the optical confines of the galaxy, with numerous point sources, while
NGC 3396 appears to be more centrally concentrated, with evidence of
some structure in the centre of the galaxy. An X-ray bridge is present
in the data, and there is evidence of point sources between the
galaxies where the disks have started to interact, as was suggested by
the HRI data.

\begin{figure*}
  \includegraphics[width=16cm]{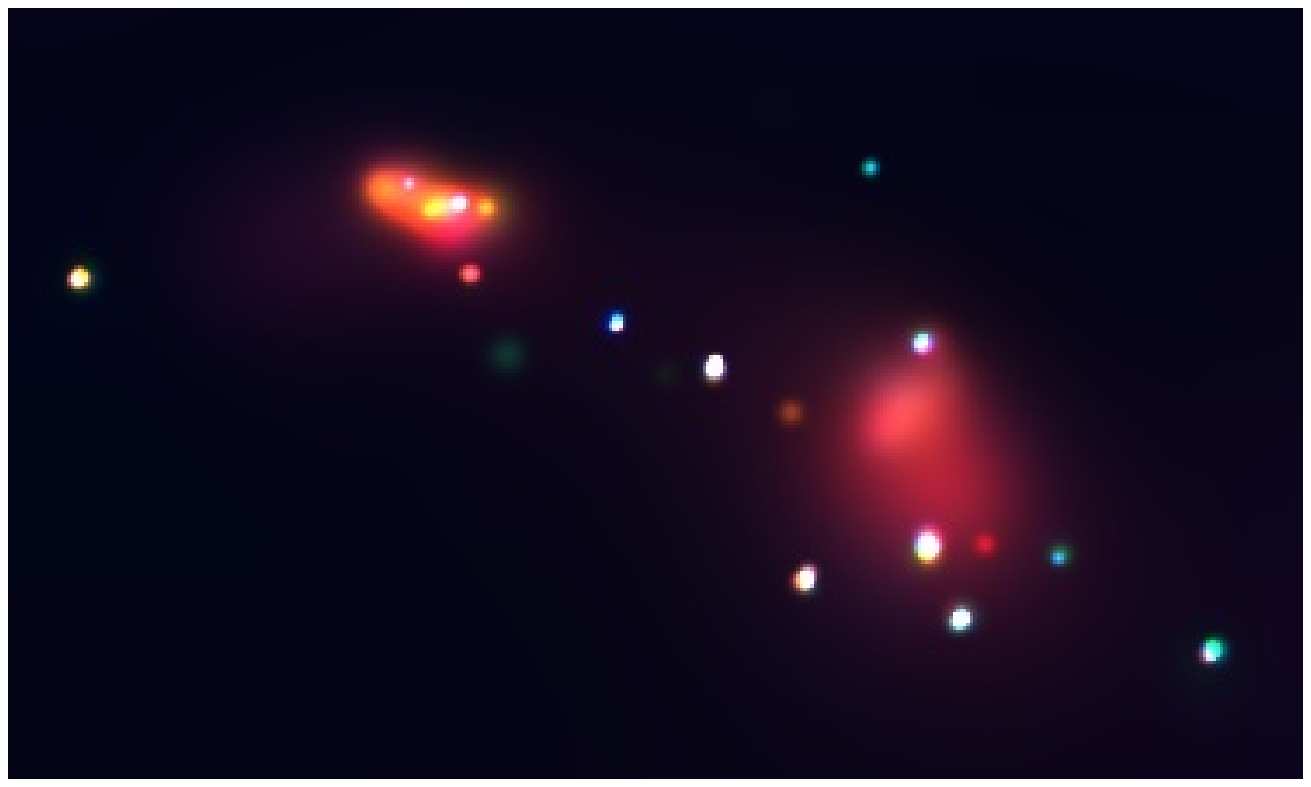}
  \caption {A `true colour' image of Arp 270. Red corresponds to 0.3$-$0.8\,keV, green to 0.8$-$2.0\,keV and blue to 2.0$-$8.0\,keV. Sources appear white as they emit across the full band.}
  \label{fig:truec}
\end{figure*}

Figure \ref{fig:truec} shows a true colour image of the system. Red
corresponds to 0.3$-$0.8\,keV, Green 0.8$-$2.0\,keV and Blue
2.0$-$8.0\,keV, hereafter referred to as soft, medium and hard
respectively. The images for all three bands were created in the same
way as described above, using the same set of smoothing scales for
each image. These three images were then combined to create a colour
image. The image shows that point-like sources tend to have harder
spectra than the diffuse emission; those appearing white imply
emission in all three energy bands. Blue and green sources very often
indicate heavy X-ray absorption. NGC 3395 shows a diffuse nucleus
whilst NGC 3396 has some hard point sources at the nucleus, surrounded
by diffuse gas.

\subsection{The Discrete Source Population}
\label{sec_source}

\begin{figure}
  \includegraphics[width=\linewidth]{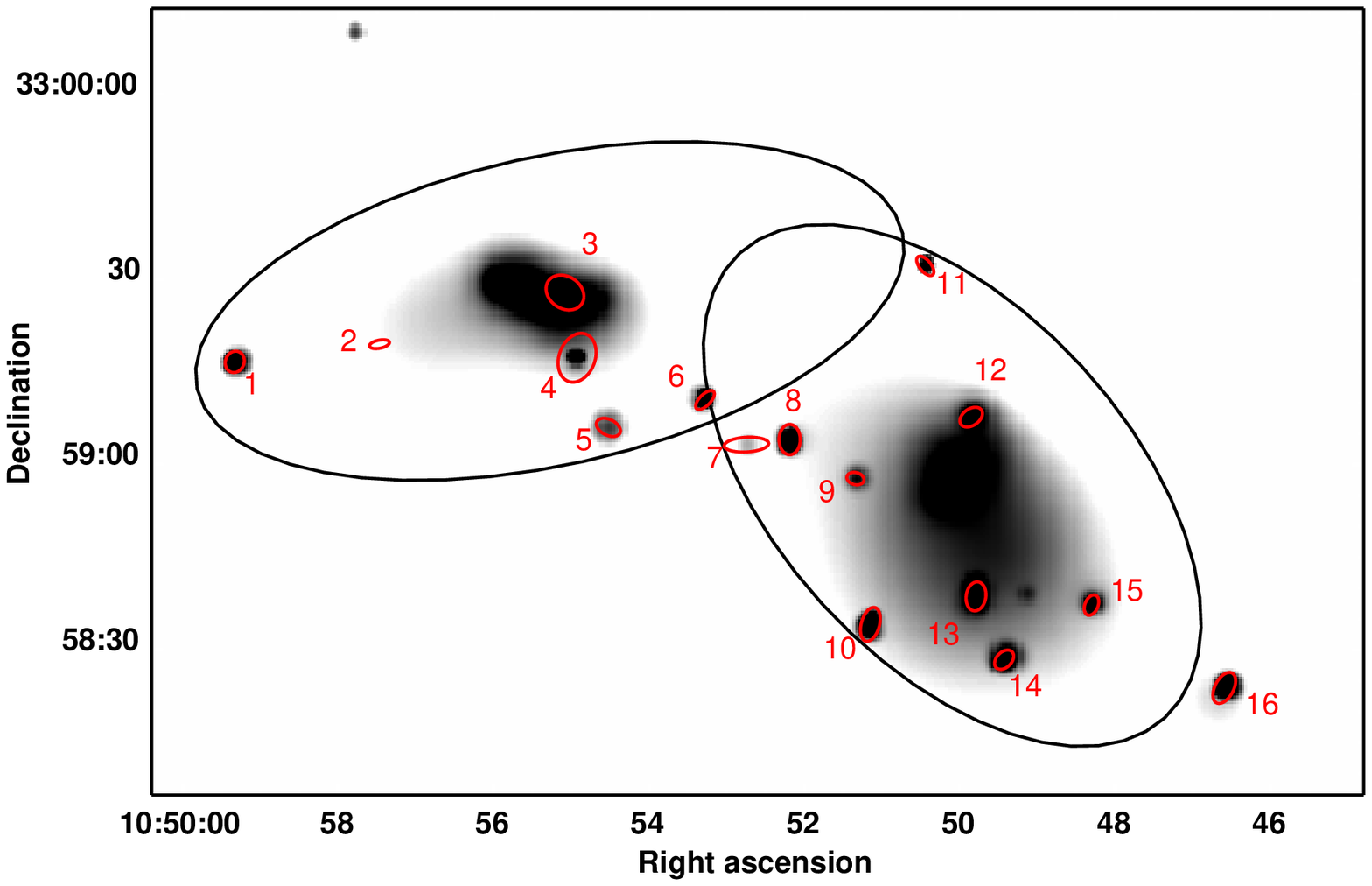} 
  \caption {The sources detected by{\em Chandra} overlaid on a grey-scale image of the X-ray emission across the full band.The extent of the D$_{25}$ ellipses is also shown. }
  \label{fig:pointsources}
\end{figure}

Discrete X-ray sources were detected using the CIAO tool {\em
wavdetect}. This was run on the full, soft, medium and hard band
images, and over the 2, 4, 8, 16 pixel wavelet scales (where pixel
width is 0.5\arcs), with a significance threshold of
$1.6\times10^{-6}$, which corresponds to one spurious source over a
800 $\times$ 800 pixel grid, the size of our image. 27 sources were
detected in the full band energy range. We then limited our sources to
those that are within the $D_{25}$ ellipses of the galaxies, where
these values were obtained from
LEDA\footnote{http://leda.univ-lyon1.fr/}. Although one of the sources
detected (specifically source 16) is seen to lie outside of the
$D_{25}$ ellipses, it is thought to be associated with an optical and
radio counterpart within the galaxies, and has therefore been included
in the subsequent analysis. The detected sources and the $D_{25}$
ellipses overlaid on the full band X-ray image are shown in Figure
\ref{fig:pointsources}, and the sources are listed in Table
\ref{tab:counts}. In this table, column (1) gives the source
identification number given in this paper, columns (2) and (3) give
the R.A and Dec. (J2000), column (4) the radius of each source region
file and column (5) the count rate for the whole observation, where an
\( ^M\) indicates {\em wavdetect} only detected a source in the
medium band. Columns
(6) and (7) indicate the count rate detected by {\em wavdetect} for
observations 1 and 2 and columns (8) and (9) give the count rate found
using aperture photometry. This was calculated by creating a region
with a radius of 1\arcs\ at the position of the source and a
background region file of a source free annulus. The number of counts
were found, which were then converted into count rates, and a
background count rate correction was applied. Of the sixteen sources coincident with the galaxy, we would expect a fraction of these to be background objects. Using the \CHANDRA\ Deep Field South number counts \citep{Giacconi_01}, we estimate the level of contamination in both the soft (0.3$-$2.0 keV) and hard (2.0$-$8.0 keV) bands. For an area comparable to the D$_{25}$ ellipses of the galaxies ($\sim$4.7 square arcminutes), an estimate of 0.45 sources in the soft band (for a flux threshold of 2.2$\times$10$^{-15}$erg s$^{-1}$) is calculated. In the hard band (with a flux threshold of 7.7$\times$10$^{-15}$erg s$^{-1}$) 0.41 background sources are estimated. This corresponds to a source contamination of 3\% in the soft band and 5\% in the hard band.
  
\begin{table*}

\centering

\caption[]{The 16 detected sources in Arp 270. Columns are explained
  in the text, an \(^M\) denotes sources detected only in the medium
  band by {\em wavdetect}.
\label{tab:counts}}
\begin{tabular}{cccccccccccc}
\noalign{\smallskip}
\hline

Source No. & R.A & Declination & Source & \multicolumn{2}{c}{Net Count Rate (count s\(^{-1} \times 10^{-3} \) )} \\
        & & & Radius & \multicolumn{3}{c}{Wavdetect} & \multicolumn{2}{c}{Aperture Photometry} \\ 
        & & & arcsec & Total Observation & Obs 1 & Obs 2 & Obs 1 & Obs 2 \\

 \hline

1  & 10 49 59.35 & +32 59 16.5 & 4.9  & 2.16 \( \pm \) 0.34       &      -                &  2.70 \( \pm \) 0.45  & 0.37 \( \pm \) 0.28  & 2.58 \( \pm \) 0.44 \\
2  & 10 49 57.50 & +32 59 19.5 & 5.4  & 0.24 \( \pm \) 0.11\(^M\) &  0.77 \( \pm \) 0.36  &          -            & 0.79 \( \pm \) 0.36  &  -                  \\
3  & 10 49 55.11 & +32 59 27.7 & 3.0  & 1.67 \( \pm \) 0.35       &       -               &  1.77 \( \pm \) 0.43  & 1.09 \( \pm \) 0.46  & 0.91 \( \pm \) 0.29 \\
4  & 10 49 54.93 & +32 59 17.3 & 4.4  & 0.98 \( \pm \) 0.26       &       -               &  0.90 \( \pm \) 0.30  & 0.37 \( \pm \) 0.28  & 0.52 \( \pm \) 0.21 \\
5  & 10 49 54.54 & +32 59 06.0 & 3.9  & 0.43 \( \pm \) 0.17       &        -              &        -              & 0.30 \( \pm \) 0.23  & 0.48 \( \pm \) 0.19 \\
6  & 10 49 53.28 & +32 59 10.2 & 3.4  & 1.58 \( \pm \) 0.30       &     -                 &  1.96 \( \pm \) 0.39  & 0.42 \( \pm \) 0.28  & 1.75 \( \pm \) 0.36 \\
7  & 10 49 52.86 & +32 59 03.3 & 3.9  & 0.40 \( \pm \) 0.15\(^M\) &      -                &        -              &                  -   & 0.47 \( \pm \) 0.19 \\
8  & 10 49 52.21 & +32 59 04.1 & 5.9  & 4.61 \( \pm \) 0.49       &  5.33 \( \pm \) 0.94  &  4.29 \( \pm \) 0.57  & 3.68 \( \pm \) 0.80  & 3.43 \( \pm \) 0.51 \\
9  & 10 49 51.35 & +32 58 57.7 & 4.4  & 0.27 \( \pm \) 0.12\(^M\) &     -                 &          -            & 0.26 \( \pm \) 0.23  & 0.41 \( \pm \) 0.17 \\
10 & 10 49 51.15 & +32 58 34.1 & 4.4  & 2.80 \( \pm \) 0.39       &  4.05 \( \pm \) 0.82  &  2.25 \( \pm \) 0.43  & 3.46 \( \pm \) 0.75  & 1.63 \( \pm \) 0.36 \\
11 & 10 49 50.45 & +32 59 32.1 & 4.9  & 0.52 \( \pm \) 0.17       &      -                &  0.71 \( \pm \) 0.24  &  -                     & 0.70 \( \pm \) 0.23 \\
12 & 10 49 49.85 & +32 59 07.8 & 5.4  & 2.20 \( \pm \) 0.36       &  1.51 \( \pm \) 0.56  &  2.41 \( \pm \) 0.45  & 0.61 \( \pm \) 0.37  & 2.16 \( \pm \) 0.40 \\
13 & 10 49 49.79 & +32 58 38.7 & 4.9  & 5.84 \( \pm \) 0.56       &  2.73 \( \pm \) 0.68  &  7.31 \( \pm \) 0.76  & 2.54 \( \pm \) 0.65  & 4.82 \( \pm \) 0.61 \\
14 & 10 49 49.43 & +32 58 28.4 & 4.9  & 3.13 \( \pm \) 0.41       &  4.40 \( \pm \) 0.85  &  2.59 \( \pm \) 0.45  & 3.97 \( \pm \) 0.80  & 1.92 \( \pm \) 0.38 \\
15 & 10 49 48.30 & +32 58 37.3 & 7.4  & 0.64 \( \pm \) 0.20       &  1.65 \( \pm \) 0.53  &      -                & 1.40 \( \pm \) 0.48  & 0.22 \( \pm \) 0.15 \\
16 & 10 49 46.60 & +32 58 23.7 & 7.4  & 2.55 \( \pm \) 0.37       &  4.27 \( \pm \) 0.84  &  1.83 \( \pm \) 0.39  & 3.99 \( \pm \) 0.81  & 1.03 \( \pm \) 0.29 \\

\noalign{\smallskip}
\hline
\end{tabular}
\end{table*}

\subsection{Spectral Analysis}
\label{sec_spec}

Spectra for the 16 detected sources were extracted from the source
region files created by {\em wavdetect}. The size of each region was
selected to ensure that as many source photons as possible were
detected whilst minimising contamination from nearby sources and
background (see source radii in Table \ref{tab:counts}). The background
files were defined as a source free annulus surrounding and concentric with each source region
file, to account for the variation of diffuse emission, and to
minimise effects related to the spatial variation of the CCD response.
Source spectra were created from both observations, using the CIAO
tool {\em psextract}.  For each source, the two spectra were fitted
simultaneously in XSPEC, tying together all the parameters, apart from
the normalisations.

Due to the short exposure time of our observations the source region
spectra have a maximum of 200 counts.  Below 50 counts, a power law
model with Galactic absorption column density along the line of sight
\NH\ = 1.98\( \times 10 ^{20} \) atom cm$^{-2}$), \footnote{calculated from
{\em nh} in FTOOLS} and photon index \( \Gamma \) = 1.5 was
assumed. Due to the low number of counts the Cash statistic \citep{Cash_79} was used
in preference to \( \chi {^2} \) when modelling the data. This was
performed in XSPEC (v11.2.0) for all 16 sources, using simple one
component models: absorbed power-law or absorbed MEKAL. The data were
restricted to 0.3$-$6.0\,keV as energies below this have calibration
uncertainties, and the spectra presented here do not have significant source flux above 6\,keV. Before the data were fitted, a correction was
applied to the auxiliary response file, to compensate for the
continuous degradation in the low energy ACIS
QE,\footnote{http://cxc.harvard.edu/ciao/why/acisqedeg.html} using the
CIAO tool {\em corrarf}.  This degradation effect is thought to be
caused by the deposition of material on the ACIS detectors or the cold
optical blocking filters. It is most prevalent at low energies and
with the continuous build up of contamination the effective low energy
QE reduces with time. The best fit models were determined by calculating the Cash statistic. These fits can be seen in
Table \ref{tab:fits}; column (1) gives the source number, column (2)
the model used, column (3) the best fit value of \NH,
column (4) \( \Gamma \) or \(T\), and columns (5) and (6) the
values for the X-ray emitted and intrinsic (i.e. corrected for
absorption) luminosities. Where statistics were too poor to constrain a multi-parameter model, the
assumed power law model described previously was fitted in XSPEC, with
all model parameters except the normalisation frozen, this is
indicated by an F in the table. A (0.3$-$6.0\,keV) flux was then
calculated from this fit.

\begin{figure}
  \vspace{0.55cm}
  \includegraphics[width=\linewidth]{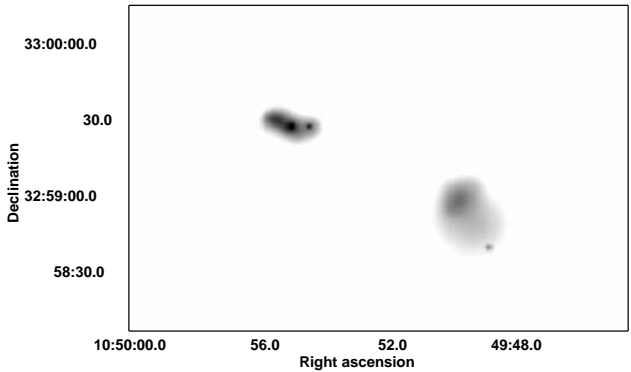}
  \caption {A grey-scale image of the smoothed diffuse gas over the full band. This image was smoothed after the point sources had been removed.}
  \label{fig:dif_gas}
\end{figure}

\begin{table*}
\centering
\caption[]{Summary of point source spectral fits. Errors on the spectral
fits parameters are given as 1\( \sigma \) for 1 interesting parameter
from XSPEC. An F denotes the value has been frozen.
\label{tab:fits}}
\begin{tabular}{cccccccccc}
\noalign{\smallskip}
\hline
Source No.  & \multicolumn{3}{c}{Spectral Fit     } & \multicolumn{2}{c}{Luminosity (0.3$-$8.0 keV)   }  \\
       & Model   & \NH & \( \Gamma / T\)
       & \multicolumn{2}{c}{(10\(^{39} \) erg s\(^{-1} \)) }    \\ 
       & & ( \( \times 10 ^{20} \)atom cm$^{-2}$) & (keV) & Observed & Intrinsic   \\      
  \\ \hline

1  & wabs*mekal & 20 \( _{-10}^{+4} \) & 2.25 \( _{-0.60}^{+0.10} \) & 0.65 \( \pm 0.08 \) & 0.97 \( \pm 0.11 \) \\
2  & wabs*power & 2  F                      & 1.5 F                       & 0.20 \( \pm 0.03 \) & 0.20 \( \pm 0.03 \) \\
3  & wabs*power & 0  \( _{-0}^{+9} \) & 0.68 \( _{-0.21}^{+0.33} \) & 1.27 \( \pm 0.17 \) & 1.27 \( \pm 0.17 \) \\
4  & wabs*mekal & 15 \( _{-6}^{+31} \) & 0.24 \( _{-0.03}^{+0.03} \) & 0.15 \( \pm 0.02 \) & 0.44 \( \pm 0.07 \) \\ 
5  & wabs*power & 2  F                      & 1.5 F                       & 0.15 \( \pm 0.03 \) & 0.15 \( \pm 0.03 \) \\
6  & wabs*power & 2  F                      & 1.5 F                       & 0.48 \( \pm 0.08 \) & 0.55 \( \pm 0.09 \) \\
7  & wabs*power & 2  F                      & 1.5 F                       & 0.21 \( \pm 0.05 \) & 0.22 \( \pm 0.05 \) \\
8  & wabs*power & 24 \( _{-6}^{+5} \) & 1.66 \( _{-0.23}^{+0.12} \) & 2.09 \( \pm 0.19 \) & 2.94 \( \pm 0.27 \) \\
9  & wabs*power & 2  F                      & 1.5 F                       & 0.10 \( \pm 0.02 \) & 0.11 \( \pm 0.02 \) \\ 
10 & wabs*power & 54 \( _{-19}^{+9} \) & 2.29 \( _{-0.95}^{+0.23} \) & 1.39 \( \pm 0.17 \) & 3.35 \( \pm 0.40 \) \\
11 & wabs*power & 2 F                      & 1.5 F                       & 0.22 \( \pm 0.04 \) & 0.23 \( \pm 0.05 \) \\
12 & wabs*power & 16 \( _{-5}^{+13} \) & 1.38 \( _{-0.91}^{+0.37} \) & 1.18 \( \pm 0.13 \) & 1.44 \( \pm 0.16 \) \\
13 & wabs*power & 41 \( _{-8}^{+5} \) & 1.88 \( _{-0.15}^{+0.15} \) & 1.92 \( \pm 0.16 \) & 3.37 \( \pm 0.29 \) \\
14 & wabs*power & 20 \( _{-14}^{+3} \) & 1.41 \( _{-0.11}^{+0.48} \) & 1.30 \( \pm 0.14 \) & 1.65 \( \pm 0.19 \) \\
15 & wabs*power & 32 \( _{-32}^{+256} \) & 1.02 \( _{-1.02}^{+1.05} \) & 0.41 \( \pm 0.05 \) & 0.50 \( \pm 0.06 \) \\
16 & wabs*power & 56 \( _{-19}^{+32} \) & 1.87 \( _{-0.18}^{+1.11} \) & 1.41 \( \pm 0.16 \) & 2.59 \( \pm 0.28 \) \\

\noalign{\smallskip}
\hline
\end{tabular}
\end{table*}

\subsection{Diffuse Emission}
\label{sec_diffuse}

From the smoothed image it is clear that there is some sub-structure
in the diffuse gas components of both NGC 3395 and NGC 3396. Once the
point sources had been identified by {\em wavdetect} they could be
removed from the data so that the diffuse emission could be seen more
clearly. We interpolated over the holes in the data using the CIAO
tool {\em
dmfilth},\footnote{http://cxc.harvard.edu/ciao/threads/diffuse\_emission}
and then smoothed using the method described previously. In Figure
\ref{fig:dif_gas}, it can be seen that there is an additional point source
in NGC 3395 in the south-west of the galaxy, and the diffuse gas is
more prevalent to the north-east. NGC 3396 has a more complex
structure in the centre of the galaxy, with the gas elongated in the
east-west direction. There also appear to be sources not formally
detected by the source-searching software.

\begin{figure*}
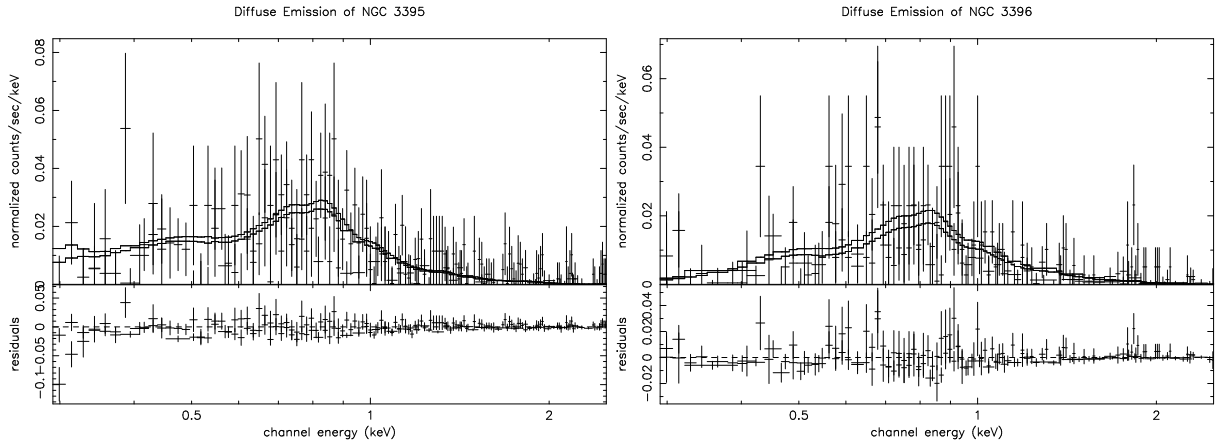

  \begin{minipage}{0.45\linewidth}
  \includegraphics[height=\linewidth,angle=270]{images/3395.ps}
  \hspace{0.5cm}
  \label{fig:spectra_5}

  \end{minipage}\vspace{0.05\linewidth}
  \begin{minipage}{0.45\linewidth}
  \includegraphics[height=\linewidth, angle=270]{images/3396.ps}
  \hspace{0.5cm}

  \end{minipage}
\caption{Diffuse spectra of the two galaxies NGC 3395 (left) and NGC
  3396 (right). The top panels in each plot show the data and the
  folded model while the bottom panels show the residuals to the
  fit. The two lines in each plot show the two data sets obtained from
  the separate pointings.
  \label{fig:spectra_6}}
\end{figure*}
We next investigated the X-ray spectra for the diffuse emission, where
the point sources identified in section \ref{sec_source} were excluded
in the region files created for each galaxy. The spectra for the two
galaxies were extracted using the CIAO tool {\em psextract} as
described previously.  Again, the low energy degradation correction,
{\em corrarf}, was applied. These data were then fitted in XSPEC using
an absorbed MEKAL model, with all the parameters tied apart from the
normalisation (see Figure \ref{fig:spectra_6}). Best fit values taken
from XSPEC are shown in Table \ref{tab:gas}. Metallicities were found
to be 0.07 \Zsol\ for NGC 3395 and 0.05 \Zsol\ for NGC 3396, but due
to the multi-temperature nature of the gas, it is likely that these
very subsolar abundances arise from fitting a single-temperature model
\citep{Buote_98} and are not a true measure of the metallicity of the
galaxies. The absorption column of NGC 3395 was found to be 4\( _{-2}^{+4} \times 10^{20}\) atom cm\(^{-2}\), which is consistent with galactic
absorption of 2 \(\times 10 ^{20} \) atom cm\(^{-2}\). The
absorption column of NGC 3396 was measured to be 9\( _{-2}^{+4} \times
10^{20}\) atom cm\(^{-2}\). This excess absorption is not unexpected
as NGC 3396 contains both HI \citep{Clemens_99} and molecular gas \citep{Smith_01,Zhu_99} and is a more compact galaxy than NGC 3395 and, as can be seen
from the HI observations \citep{Clemens_99}, is highly inclined to the
line of sight. 

\begin{table*}
\begin{center}
\caption{Best results from fitting MEKAL model to the spectra of the diffuse gas components of NGC 3395 and NGC 3396
\label{tab:gas}}
\begin{tabular}{cccccc}
\noalign{\smallskip}
\hline
Galaxy  & \NH              & $T$    & \Z           & \multicolumn{2}{c}{Luminosity}         \\
        & (\( 10^{20} \) atom cm$^{-2})$         & keV   & \Zsol           & \multicolumn{2}{c}{(10\(^{39} \) erg s\(^{-1} \)) } \\
        &  &       &  & Observed & Intrinsic \\ 

 \hline

NGC 3395  & 4 \( _{-2}^{+2} \) & 0.52 \( _{-0.10}^{+0.05} \) & 0.07 \( _{-0.01}^{+0.03} \) & 3.50 \( \pm \) 0.12 & 4.64 \( \pm \) 0.16 \\
NGC 3396  & 9 \( _{-3}^{+18} \) & 0.49 \( _{-0.09}^{+0.05} \) & 0.05 \( _{-0.01}^{+0.04} \) & 2.86 \( \pm \) 0.14 & 4.87 \( \pm \) 0.24 \\
\noalign{\smallskip}
\hline
\end{tabular}

\end{center}
\end{table*}

\subsection{X-ray Variability}
\label{sec_varib}

\begin{figure*}
  \includegraphics[width=14cm]{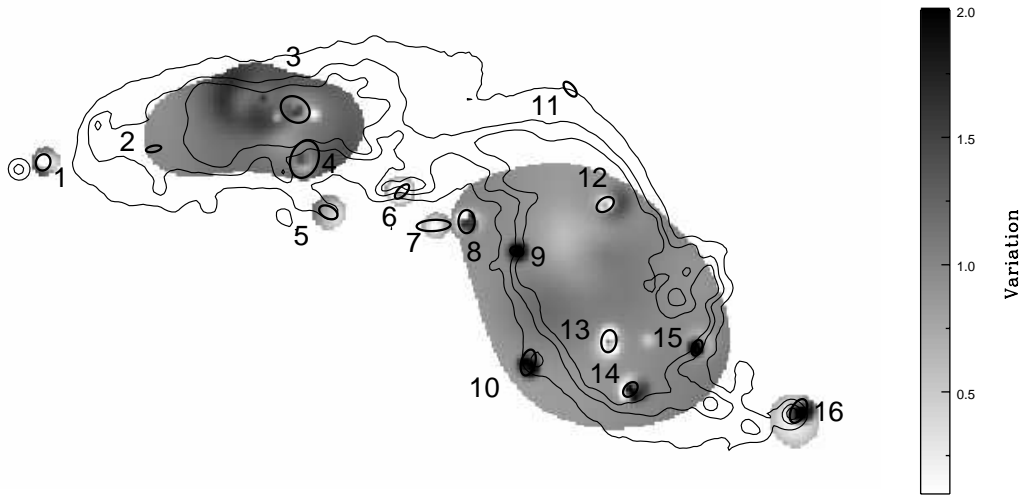}
  \hspace{0.5cm}

 \caption {A Variation Map of the two observations, the lighter
 features indicate a higher flux in the second observation. Optical contours are overlaid, and numbered ellipses denote sources detected by {\em wavdetect}. }
  \label{fig:vmap}
\end{figure*}

Our observation was taken in two separate pointings, 6 months
apart. Because of this we are able to examine the variability within
the system. Sources have been defined to be variable if the difference
in count rates, given in columns (8) and (9) of Table \ref{tab:counts},
is greater than the 2\( \sigma \) count rate error for the two
observations, 
\begin{equation}
\Delta \mathrm {flux} > 2\sqrt{(\sigma_1^2 + \sigma_2^2)}.
\end{equation}
Sources which have been found to vary are 1, 6, 10, 13, 14, 15 and 16, of
which sources 10, 13, 14 and 16 are classed as {\em Ultraluminous X-ray
Sources} (ULX's), with \LX\(\ge 10 ^{39} \)erg s\(^{-1}\).  A
variability map has been created so that the locations of the variable
sources can be more readily identified (Figure \ref{fig:vmap}). The
map was created by taking a ratio of the two observations, the darker
features on the map indicate a higher flux in the first
observation. This map is a useful tool in highlighting any regions of
unresolved point sources. In NGC 3395 a point-like feature can be seen
to the west of point source 13. This feature appears lighter than the
diffuse gas indicating that there was a greater flux during the second
observation. NGC 3396 shows substructure in the centre of the galaxy,
with some darker and lighter features indicating that there are point
sources not detected by the source-searching algorithms.

\section{Discussion}
\label{sec_disc}

\subsection{Source Population}

The \CHANDRA\ ACIS-S has spatial resolution (FWHM) of 0.5\arcs, and given the
distance to Arp 270 of 28 \Mpc , one can resolve point sources
separated by $\sim$70 pc. This has enabled more point sources to be
detected than were previously found with the \ROSAT\ HRI, as can be
seen by comparing Table \ref{tab:hri} with Table \ref{tab:counts}. Of
the four sources detected by the HRI, sources H1 and H3 correspond
to source 3 and source 16 in the \CHANDRA\ data, source H2 lies
within 2\arcs\ of source 12, and source H4 was not included
in the \CHANDRA\ data presented here, although it was detected, as it lies $\sim$1\arcm\ outside
the D$_{25}$ ellipses. 

\subsubsection{Spectral trends}

\begin{figure*}
  \begin{minipage}{0.44\linewidth}
  \centering
  
    \includegraphics[width=\linewidth]{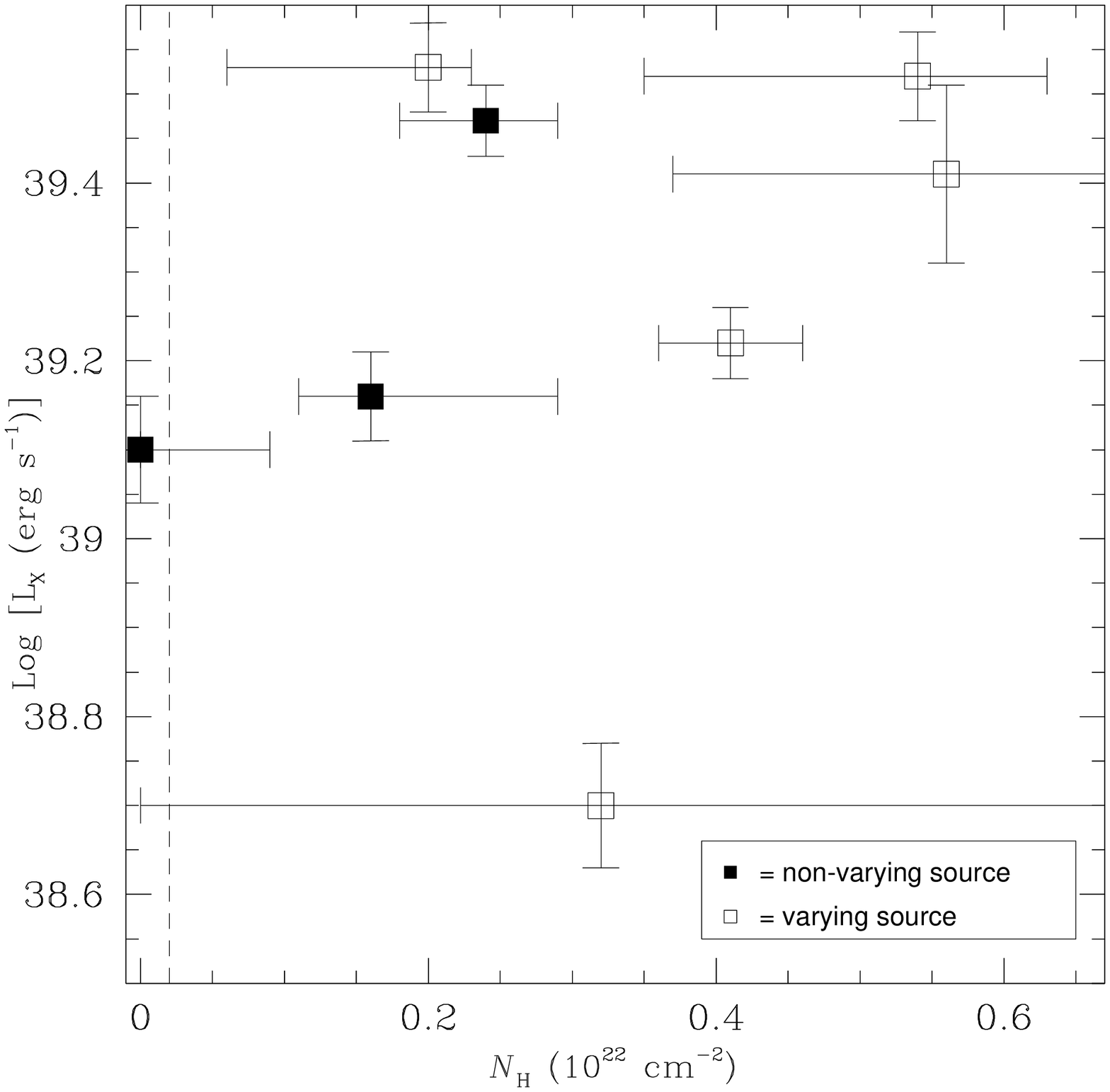}
    \caption{A plot of luminosity against column density,  for 8 of the 16 point sources, the other 8 have not been plotted due to either the low number of counts or they were fitted with an alternative model (see text), the dashed line indicates the Galactic line of sight column density.}
    \label{fig:lum_nh}
  
  \end{minipage}\hspace{0.02\linewidth}
  \begin{minipage}{0.44\linewidth}
  \centering

    \includegraphics[width=\linewidth]{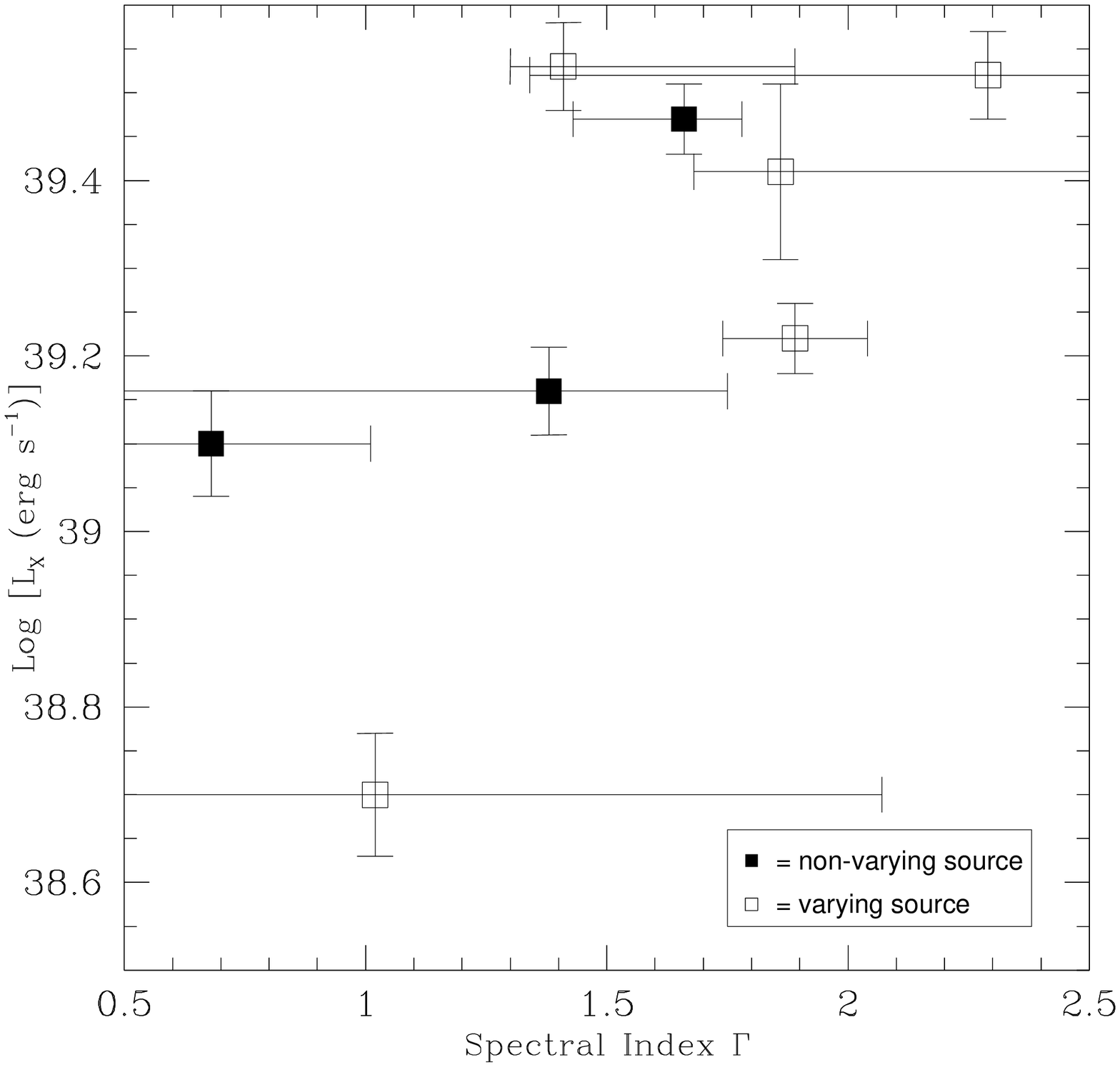}
    \caption{As for Figure \ref{fig:lum_nh}, plotting luminosity against the photon index \(\Gamma \).    }
    \label{fig:lum_gam}

\end{minipage}

\begin{minipage}{0.45\linewidth}
  \centering

    \includegraphics[width=\linewidth]{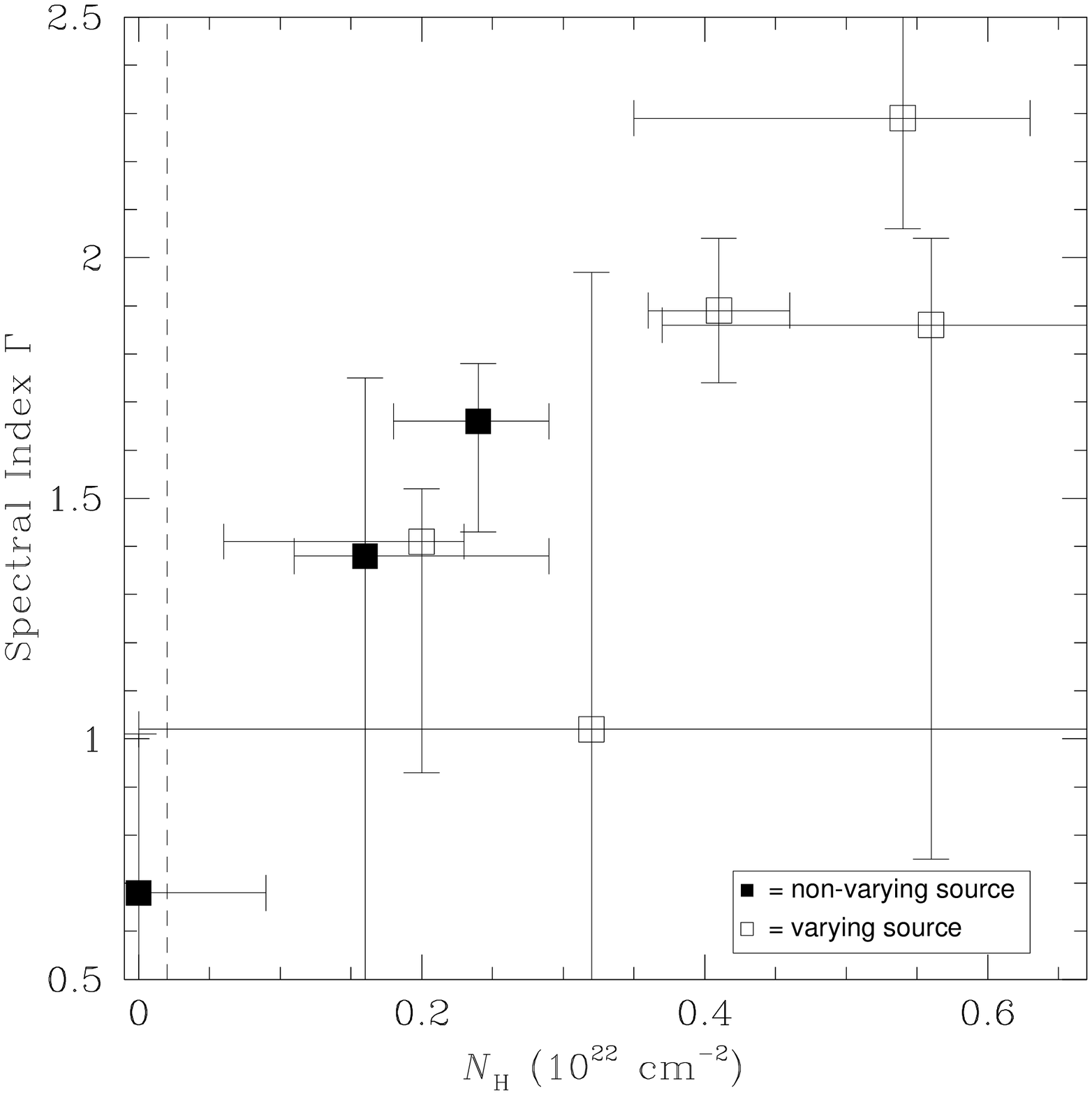}
    \caption{As for Figure \ref{fig:lum_nh}, plotting \(\Gamma \) against column density, the dashed line indicates the Galactic line of sight column density. }
    \label{fig:nh_gam}

\end{minipage}
\end{figure*}

From Table \ref{tab:fits}, it can be seen that there is a wide range of
spectral parameters within the source population of Arp 270.  It is
useful to investigate the spectral distribution of the point sources
to establish if there are any spectral trends, or if the variable
sources have different properties from the other sources. In Figure
\ref{fig:lum_nh} the intrinsic luminosity is plotted against the best fit value
of \NH. Figure \ref{fig:lum_gam} plots intrinsic
luminosity against the best fit power law index (\( \Gamma \)), and
Figure \ref{fig:nh_gam} shows \( \Gamma \) against \NH. In these
three figures only 8 of the 16 detected sources have been plotted. Of the eight omitted sources, six have low counts, such that the model parameters cannot be well constrained, and the remaining two sources are best fitted with a MEKAL model. The 8 sources that have been plotted have reliable error bounds for \NH, \( \Gamma \) and intrinsic luminosity.

Trends were identified using the Kendall's rank coefficient which corresponds
to a correlation significance in units of Gaussian sigma.  Applying this test finds no strong correlation between the properties of the point sources and their variability. However, there does appear to be a weak trend for sources with higher values of \NH\ to show variability. This could be due to sources with higher values of \NH\ being close to the galactic plane, indicating that they are likely to be young objects and as such we would expect to see more variability. Alternatively, it could be that this absorption is intrinsic and variable, leading to an observed change in luminosity for these sources. The plot of X-ray luminosity against \NH\ (Figure \ref{fig:lum_nh}) shows no correlation between the intrinsic luminosity and the amount of
absorption, but there does appear to be a trend (1.48$\sigma$) for more
luminous sources to have softer spectra (Figure
\ref{fig:lum_gam}). Also, there is a clear correlation of 2.23$\sigma$ between
the photon index and the amount of absorption. This trend is primarily
driven by a population of XRBs, which have both a high
absorbing column and a soft X-ray spectrum (Figure
\ref{fig:nh_gam}), indicating that they could be black hole
binaries (BHBs). Even though these sources show a tendency for
increasing softness with luminosity, their spectra would still classify them as
sources in a low/hard state. This state is characterised by power law spectral indices in the range 1.5 $<$\(\Gamma\) $<$2.0, and
an accretion rate at a low fraction of the Eddington luminosity of the
compact object \citep{Tanaka_95,White_95}. But, this definition would
require objects with a mass of $\sim$10$-$400\Msol\ for the
luminosities we have observed. This would range from a stellar-mass black hole ($M\sim$3$-$20\Msol\,) to an intermediate-mass black hole (IMBH; $M\sim10^2-10^4$ \Msol\,) \citep{Miller_03}. However, it has recently been suggested that the states of BHBs cannot
be defined by their luminosity \citep{McClintock_03} as it is now
clear that the spectral hardness of a BHB is not a direct function of mass
accretion rate \citep{Homan_01}. In fact, giving support to this, high
luminosity, spectrally hard sources have been detected in NGC
4485/4490 \citep{Roberts_02} and the transition from a low/soft
spectral state to a high/hard one has been observed in some ULXs in
The Antennae \citep{Fabbiano_03}, the ULX in NGC 7714 \citep{Soria_04} and the ULX in Holmberg II X-1
\citep{Dewangan_04}. Many of the discrete sources in Arp 270 have high values of \NH, indicating that they could lie within, either a gas
rich environment, or in close proximity to the galactic
plane. Given the location of these sources, it seems likely that they were triggered in the last episode of star formation. Therefore, we feel it is probable that the sources we have
identified are high mass X-ray binaries (HMXRBs), due to both the
nature of the individual sources and the system they lie within.
\begin{figure*}
  \begin{minipage}{0.45\linewidth}
  \includegraphics[height=\linewidth]{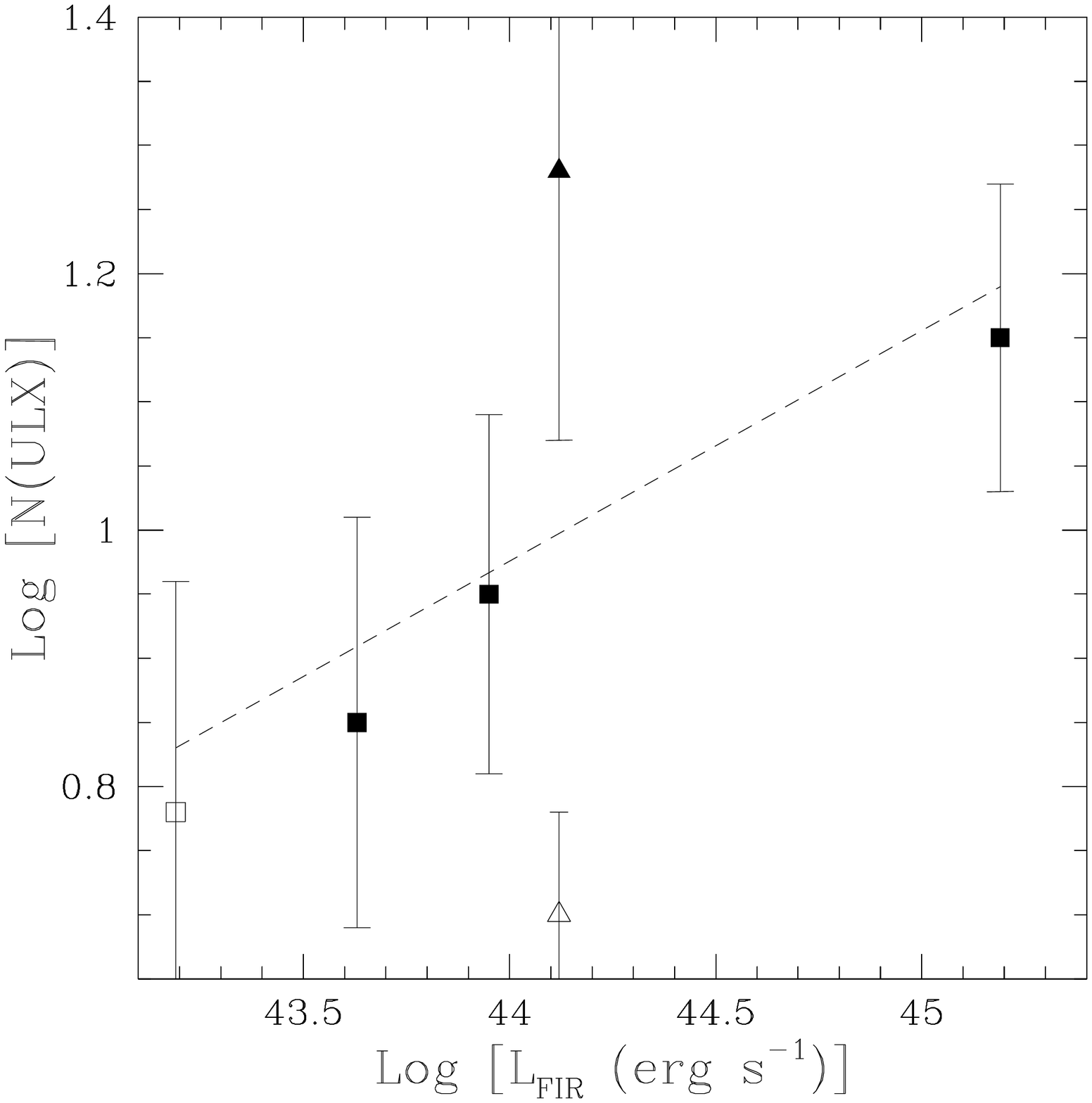}
  \hspace{0.1cm}

  \end{minipage}\vspace{0.05\linewidth}
  \begin{minipage}{0.45\linewidth}
  \includegraphics[height=\linewidth]{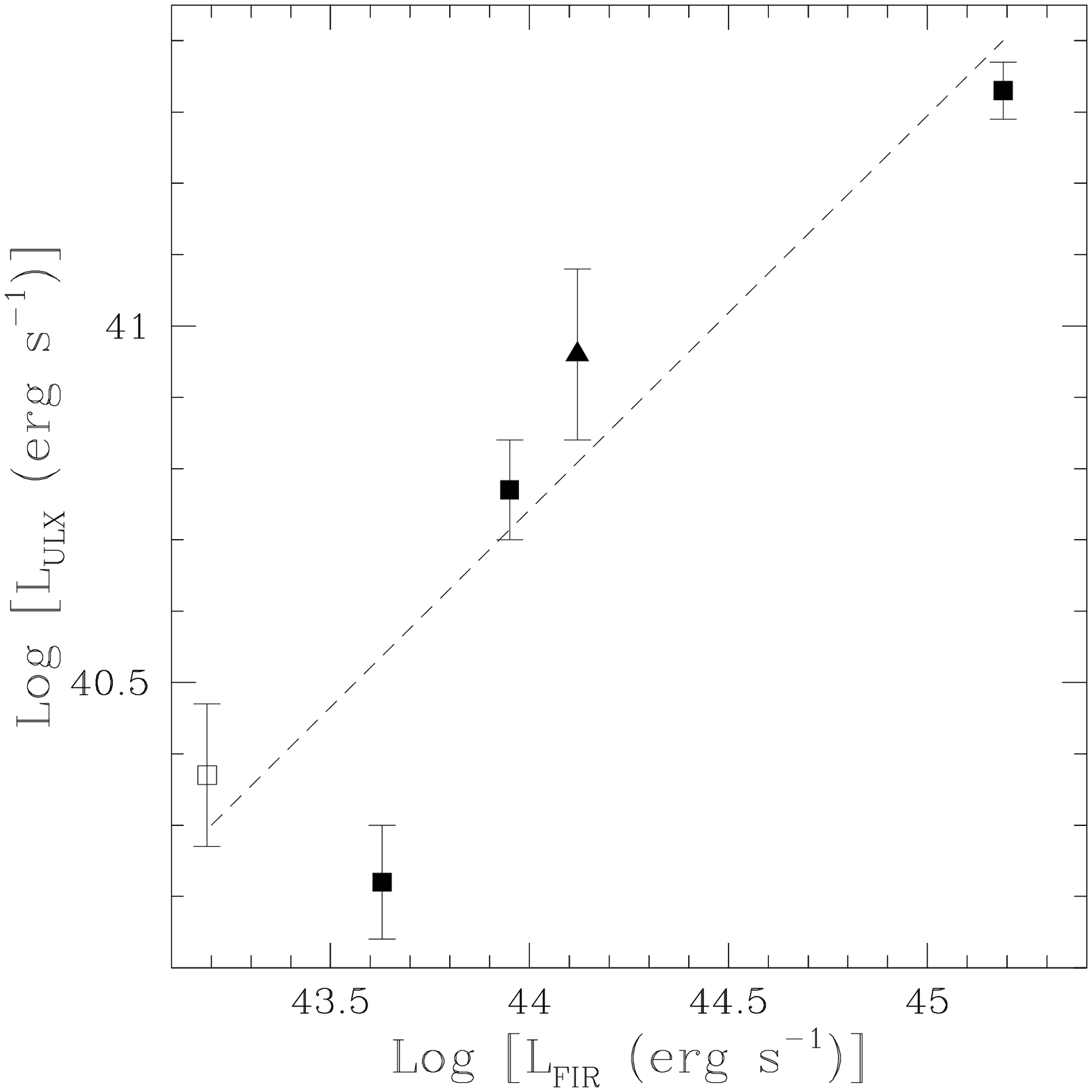}
  \hspace{0.1cm}

  \end{minipage}
  \caption{Left shows a comparison of \NULX\ per galaxy against \LFIR, a fit to the data gives a slope of 0.18$\pm$0.13. The open square indicates the number of ULXs in NGC 4485/90, this system, unlike the rest of the sample, has a small mass ratio (see text). The open triangle indicates the number of ULXs in The Mice above a source detection limit of \LX$= 5\times 10^{39}$erg s$^{-1}$, whilst the filled triangle indicates the corresponding estimated number for \LX $\ge 1\times 10^{39}$erg s$^{-1}$. Right shows \LULX\ against \LFIR\ with a slope of 0.54$\pm$0.04, the open square indicates \LULX\ for NGC 4485/90 and the triangle indicates the estimated value of \LULX\ for The Mice, derived from the universal LF from \citet{Grimm_03}.
  \label{fig:LFIR}}

\end{figure*}
Out of the sixteen sources detected by {\em wavdetect}, seven exceed
the Eddington limit for a 7\Msol\ star of 1\( \times 10^{39} \) erg s$^{-1}$ and are therefore classified as ULX sources. The spatial
resolution of these sources has been investigated to see if there is
any evidence of extension. This was done by creating radial profiles
of each source, generating PSF models with the CIAO tool {\em mkspf},
and comparing in {\em ChIPS}. From this it can be concluded that all
of the 16 sources are point-like.

\subsubsection{The relationship between ULXs and \LFIR}

In a recent paper by \citet{Swartz_04}, over 80 galaxies imaged by
\CHANDRA\ were analysed for ULXs. From their analysis a correlation
between the far infrared luminosity (\LFIR), an indicator for star
formation, and ULXs in interacting galaxies was found. There is also
evidence that the number of ULXs (\NULX) and the level of \LFIR\
significantly increases with merging and interacting galaxies.  To
investigate this correlation we have compared Arp 270 to four other
interacting galaxy pairs (Table \ref{tab:ulx}). These pairs were selected such that they have
been studied with \CHANDRA\
\citep{Fabbiano_01,Lira_02,Roberts_02,Read_03} and that the two
galaxies are in a pre-merger stage. We attempt to compare systems comparable to Arp 270, although we note that one of the systems, NGC 4485/90, has a mass ratio of 1:4.4,\footnote{Taken from the Two Micron All Sky Survey (2MASS); http://irsa.ipac.caltech.edu/Missions/2mass.html} not $\sim$1:1 as the other systems have. Out of the four systems, three are thought to be at more advanced stages of merging
than Arp 270; The Mice \citep{Read_03}, The Antennae
\citep{Fabbiano_01} and NGC 3256 \citep{Lira_02}. Comparison of N-body simulations of NGC 4485/90 \citep{Elmegreen_98} with Arp 270 suggests that NGC 4485/90 is at an earlier stage of evolution than Arp 270. 
\begin{table*}
\begin{center}
\caption{A comparison of the number of ULXs (\LX $\ge10^{39}$erg s$^{-1}$) in a system and its
\LFIR, an indication of star formation.  $^\mathrm{a}$ indicates system has a source detection limit of \LX$ =5\times 10^{39}$erg s$^{-1}$.}
\label{tab:ulx}

\begin{tabular}{c@{}cccc@{}cc}   \\

 \noalign{\smallskip}
 \hline
Galaxy & No of ULXs & Log \LFIR  & \LFIR/\LB & Mass Ratio & Mass Ratio Reference & Type \\
System &            & (erg s$^{-1}$)&        &            &                      &      \\
\hline 

NGC 4485/90  & 6       & 43.19 & 0.55 & 1:4.4$^{\beta}$      & 2MASS                & 1$^\mathrm{st}$ Approach  \\
Arp 270      & 7       & 43.63 & 0.81 & 1:1.3$^{\gamma}$     & \citet{Hern_01}      & 2$^\mathrm{nd}$ Approach  \\
The Mice     & 5$^\mathrm{a}$ & 44.12 & 1.13 & 1:1.3$^{\beta}$ & \citet{Gavazzi_96} & Disks Colliding   \\
The Antennae & 9       & 43.95 & 0.94 & 1:1.05$^{\gamma}$    & \citet{Lauberts_89}  & Disks Colliding   \\
NGC 3256     & 14      & 45.19 & 5.83 &  $^\mathrm{f}$       & -                    & Near nuclear coalescence  \\

\noalign{\smallskip}
\hline

\end{tabular}

\end{center}
Notes: $^\beta$Calculated from the R band magnitudes, $^{\gamma}$ calculated from the K band magnitudes. $^\mathrm{f}$ individual luminosities for the two galaxies cannot be determined but from the N-body simulations of the system \citep{English_03} and the existence of the long tidal tails it is likely that this is the product of a near-equal-mass-merger.
\end{table*}
The correlation between \LFIR\ and \NULX\ is investigated in Table
\ref{tab:ulx}, along with a measure for star formation activity,
\LFIR/\LB\ and the merger status of each system. Column (1) gives
the system name, column (2) \NULX\ detected in that system,
column (3) Log \LFIR, (4) \LFIR/\LB, column (5) the mass ratio and column (6) the
status of the system.  The FIR luminosities are calculated using the
expression \citep{Devereux_89}
\begin{equation}
\label{eq:lfir}
\LFIR =3.65 \times 10^5[2.58S_{60 \mu m}+S_{100\mu m}]D^2\Lsol,
\end{equation}
with {\em IRAS} 60- and 100-$\mu$m fluxes taken from the {\em IRAS}
Point Source Catalogue \citep{Moshir_90}. The optical (B) luminosities
were calculated as in \citet{Tully_88}
\begin{equation}
\mathrm{log} \LB\, (\Lsol) =12.192-0.4B_\mathrm{T}+2\mathrm{log}D,
\end{equation}
where $B_\mathrm{T}$ is the blue apparent magnitude and $D$ is the
distance in Mpc. Values of blue apparent magnitude were taken from
\citet{Dev_91} (the value for The Mice was taken from NGC 2000.0
\citep{Dreyer_88}). From this table it can be seen that \LFIR/\LB\ correlates with the stage of evolution that the system is thought to be in, with
NGC 4485/90 and Arp 270 having lower values of \LFIR/\LB\ than The Mice
and The Antennae, which in turn have a lower value of \LFIR/\LB\ than
NGC 3256. 

We find a weak correlation when we compare \LFIR\ and
\NULX\ per galaxy (0.94$\sigma$); from Figure \ref{fig:LFIR} it can be seen that The Mice is an outlier from this trend. However it is quite feasible that the smaller number of ULXs detected in The Mice is caused by the higher source detection limit of \LX $=5\times 10^{39}$erg s$^{-1}$ for this observation \citep{Read_03}. To extrapolate \NULX\ to \LX $\ge 1\times 10^{39}$erg s$^{-1}$ for The Mice, we used the universal luminosity function derived by \citet{Grimm_03}. When this value is used to compare \LFIR\ and
\NULX\ per galaxy (Figure \ref{fig:LFIR}) a stronger correlation of 1.96$\sigma$ is found. Plotting log \NULX\ against log \LFIR\ gives a slope of 0.18$\pm$0.13 (excluding NGC 4485/90 from this fit we find a slope of 0.18$\pm$0.11 with a correlation of 1.36$\sigma$). We also found a correlation (1.96$\sigma$) between the luminosity of the ULXs (\LULX) and \LFIR, the value of \LULX\ for The Mice was again extrapolated from the universal LF of \citet{Grimm_03}. Fitting log \LULX\ against log \LFIR\ gives a slope of 0.54$\pm$0.04 (This changes to a slope of 0.57$\pm0.04$ with a correlation of 2.04$\sigma$ when excluding NGC 4485/90)(Figure \ref{fig:LFIR}). \citet{Read_01} found the total point source X-ray luminosity for a sample of both normal and starburst galaxies to be correlated with \LFIR\ with a slope of (0.66$\pm$0.09), which is in agreement, within errors, with our fit. In \citet{Colbert_03} an expression to estimate the point source X-ray luminosity ($L_\mathrm{XP}$) was derived. 
\begin{equation}
L_\mathrm{XP}=\alpha L_\mathrm{K}+\beta L_\mathrm{FIR+UV}.
\end{equation}
$L_\mathrm{XP}$ is made up of contributions from both old and young stellar populations, which correlate with the K-band and FIR+UV luminosities respectively. These correlations motivate the relation derived in \citet{Colbert_03}. The contribution from the old population contains some lower luminosity ULXs \citep{Jeltema_03} while the young population, connected with SFR, contains the most luminous ULXs \citep{Colbert_03,Swartz_04,Wolter_04}. Our results give further credence to this finding, with our \LULX\ against \LFIR\ fit having a steeper slope than our \NULX\ against \LFIR\ fit, indicating that the systems with higher \LFIR\ contain more luminous ULXs. This suggests that ULXs are a heterogeneous class, made up from a contribution of old and young stellar populations.

\subsubsection{Correlation with UV emission}

\begin{figure*}
  \centering

    \includegraphics[width=12.35cm]{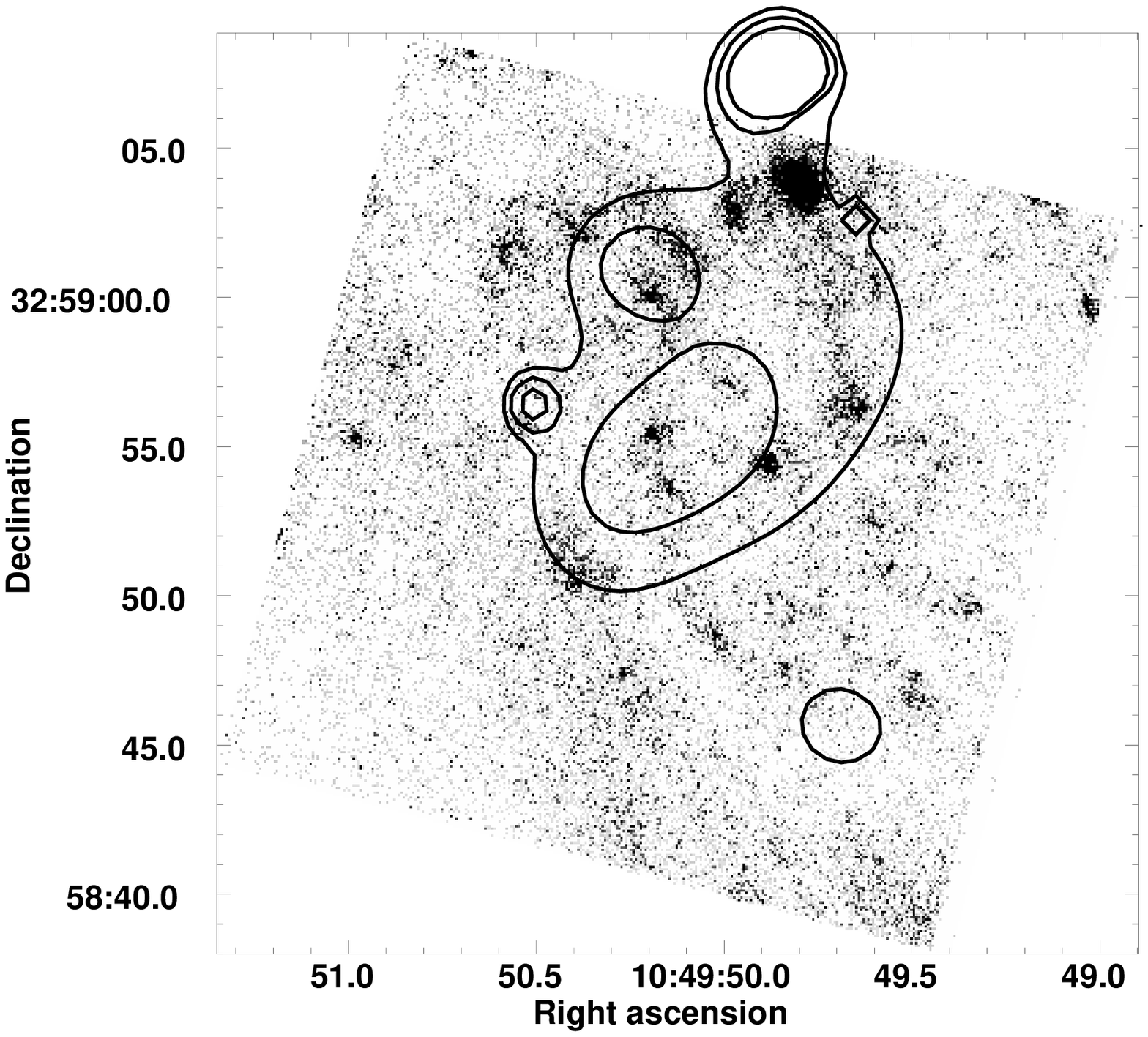}
    \hspace{6mm}

  \centering

    \includegraphics[width=12cm]{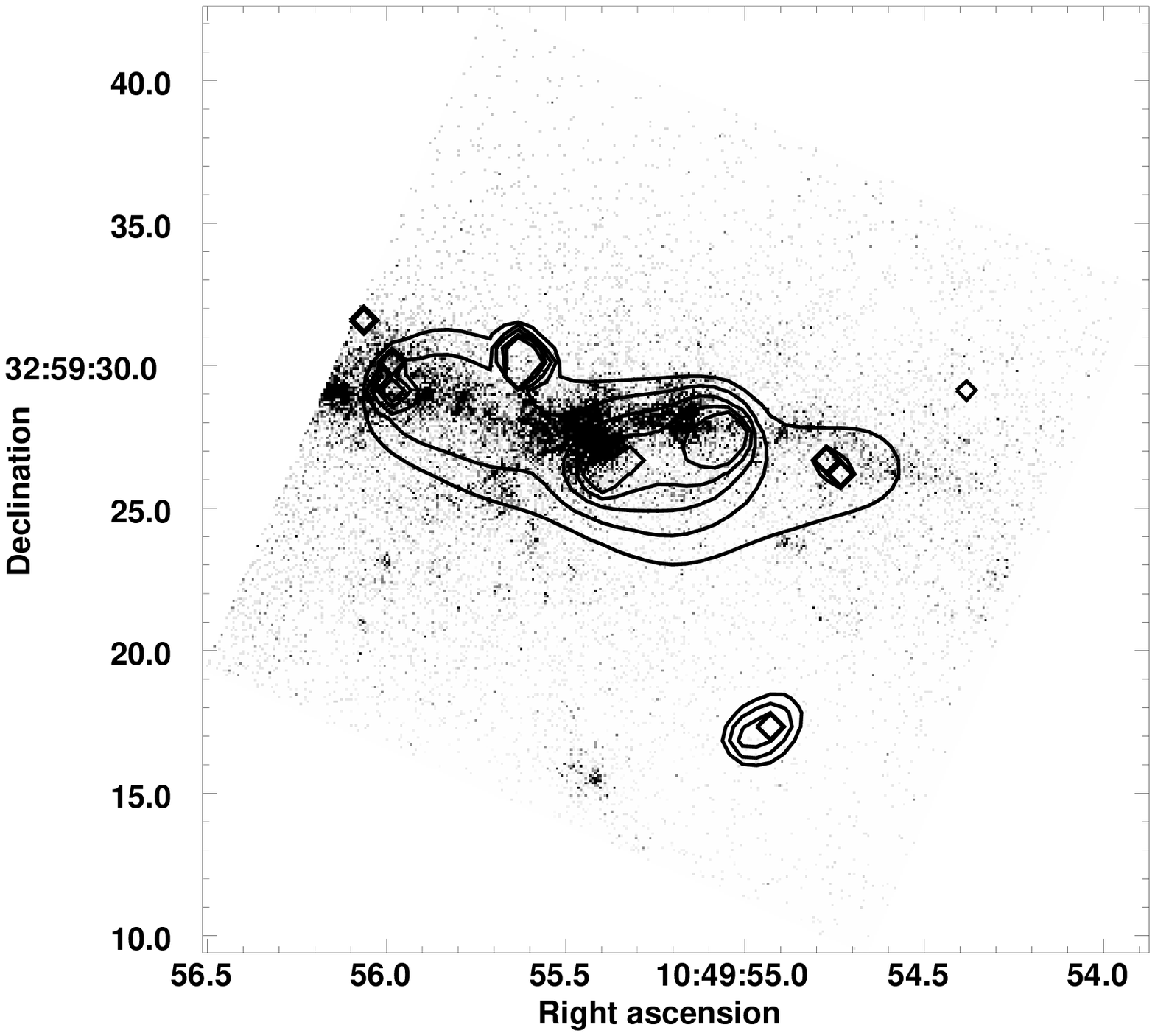}
    \caption{Adaptively smoothed 0.3 - 8.0\,keV X-ray contours overlaid
   on STIS FUV images of the central regions of NGC 3395 (top) and NGC 3396
   (bottom). Contours increase by a factor of $\sqrt{2}$. Note that the diamond features in the images (e.g. the isolated diamond to the right in the bottom figure) are low level contours, not marked features.}
    \label{fig:fuv}

\end{figure*}

\citet{Immler_03}, in a study of the late-type spiral galaxy NGC 1637, have observed that $\sim$15\% of X-ray point sources in this system are associated with young globular clusters, or star forming knots. Further associations between luminous X-ray point sources and optical counterparts have also been observed in starburst and interacting galaxies, as well as in late-type spirals \citep{Kaaret_04,Liu_04,Smith_04,Soria_05}. The central regions of both galaxies in Arp 270 are shown in Figure \ref{fig:fuv} where STIS (Space Telescope Imaging
Spectrograph) far UV images from the {\em Hubble Space Telescope}
\citep{Hancock_03} are shown with X-ray contour overlays of the
\CHANDRA\ full band emission for both NGC 3395 and NGC 3396. The X-ray
contours for both galaxies were created from smoothed images, with
lower values of smoothing than were previously used. This was done so
that the features of both galaxies could be more clearly seen.
 From \citet{Hancock_03}, one gravitationally bound knot was identified
in NGC 3395, and six in NGC 3396. The only formally identified X-ray
source to have a UV counterpart is source 3 (see Figure \ref{fig:pointsources} for source numbering), but as previously
mentioned, both galaxies have substructure which is more clearly seen
in Figure \ref{fig:fuv}, due to the reduced smoothing. Even so, many
of the knots are not detected in the X-ray, implying that they have luminosity below our threshold of 1.4 $\times$ 10$^{38}$ erg s$^{-1}$.  X-ray sources with optical counterparts (from the APM Catalogue\footnote{http://www.ast.cam.ac.uk/$\sim$apmcat/}) were used to check
the astrometry, the absolute X-ray source positions are accurate to better than 0.5\arcs. In NGC 3396 there are two X-ray sources that have been found to
be offset from star forming knots in the UV data by 1.0\arcs\ and 1.5\arcs. The second is
source 3, while the first has not been formally identified by {\em
  wavdetect}  (these can be seen in the bottom plot in
Figure \ref{fig:fuv}). These offsets correspond to displacements of
140 \pc\ and 210 \pc\ respectively. Given that there are 6 UV knots and 4 X-ray sources within 11.6 kpc$^2$ (the area of the UV plate, which lies within the D$_{25}$ ellipse of NGC 3396), the probability of finding one UV source within 140 pc of an X-ray source is 0.12, and the probability of finding two UV sources within 210 pc of X-ray sources is 0.10. This indicates that these associations are moderately unlikely to be chance near-alignments.

It may be that the offsets indicate that these sources
are runaway binaries, HMXRBs that have been kicked out by an
asymmetric explosion when the neutron star forms \citep{Lyne_94}. Taking the
mass of both these knots as $M\geq$ 2$\times 10 ^{6}$\Msol\ \citep{Hancock_03},
and assuming that the kickout velocity is equal to the escape velocity for
the knot, we calculate $V_\mathrm{esc} \geq $ 60 km s$^{-1}$ which corresponds to
knot ages of $\leq$ 2.3 Myr and $\leq$ 3.5 Myr, respectively. Previous studies have found kickout velocities of
30$-$100 km s$^{-1}$ \citep{Cordes_98, Fryer_01}. Using these values we
find the first knot to be between $\sim$1.5 Myr and
$\sim$5 Myr and the second knot ranges from $\sim$2
$\times$ Myr and $\sim$10 Myr in age. These
ages are in agreement with the values found by \citet{Hancock_03},
who find evidence for star forming knots as young as
5 Myr. The spectral hardness and luminosity of source 3, coupled with the source's close proximity to the centre of the galaxy,
suggests the possibility that it might be a low luminosity AGN. However, from
\citet{Clemens_99} it can be seen that the dynamical centre of the galaxy
is offset from this source by $\sim$3\arcsec, making this interpretation unlikely. 
%
%

\subsubsection{Correlation with \Halpha\ and radio emission}

\begin{figure}
  \includegraphics[width=\linewidth]{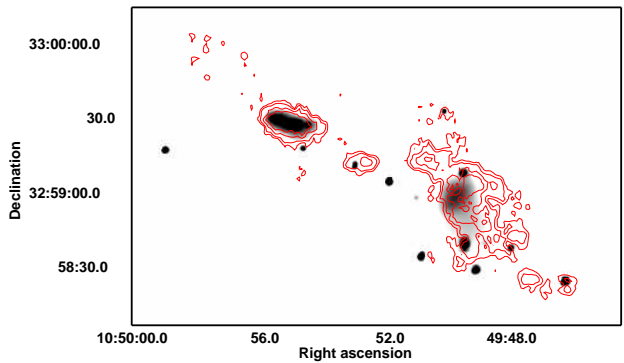}
  \hspace{0.5cm}
 
  \caption {Contours of \Halpha\ increasing by a factor of
  $\sqrt3$ from \citet{Garrido_02} overlaid on 0.2 - 8.0 keV X-ray
  emission.}
 \label{fig:halpha}
\end{figure}

Figure \ref{fig:halpha} shows \Halpha\ emission contours
 \citep{Garrido_02} overlaid on the smoothed, full band, X-ray
 emission. It can be seen from this figure that the position and morphology of the \Halpha\ emission closely follows the X-ray emission at the centres of the two galaxies, indicating that these are regions of active star formation. In particular, the good
 spatial correlation seen between several compact sources detected in
 both X-ray and \Halpha\ are good indicators of the presence of rich
 associations of massive OB stars and therefore HMXRBs. One
 association of particular note is with source 6 (see Figure \ref{fig:pointsources} for source numbering), which lies between the
 two galaxies. It is probable that star formation in this region is taking place now, as a direct consequence of the collision of the two disks of the galaxies. The \Halpha\ emission from NGC 3395 is more diffuse than that in NGC
 3396, which agrees well with the UV star forming knots for both
 galaxies. NGC 3395 has 66 identified knots \citep{Hancock_03} but
 only one that is thought to be gravitationally bound and could
 therefore evolve into a globular cluster,  whereas, NGC 3396 has 51
 identified knots with 6 potential globular clusters. Consequently the
 \Halpha\ emission will appear more diffuse in NGC 3395 as the
 regions of star formation are smaller and more dispersed than in NGC 3396.

\begin{figure}
  \includegraphics[width=\linewidth]{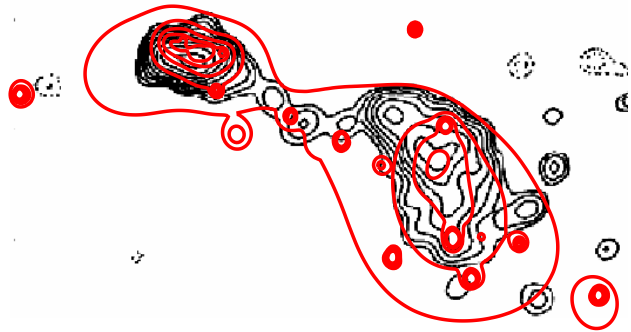}
  \hspace{0.5cm}

 \caption {Radio map at 8415 MHz (thin, black contours), with 0.3-8 \,keV X-ray contours
 overlaid (thick, red contours). Radio contours are taken from \citet{Huang_94}.}
  \label{fig:rad_high}
\end{figure}

Further evidence of recent star formation in the centres of both
galaxies is presented in the radio observation of the
galaxy pair by \citet{Huang_94}. 
%
Figure \ref{fig:rad_high}
shows 8415 MHz radio contours with full band X-ray contours overlaid. From this figure a radio continuum bridge between the two galaxies can be seen. This bridge has also been detected optically in the studies of
\citet{Carpenter_57}, with the optical and radio emission in good
agreement. The optical knots in the bridge
suggest current or recent star formation and the radio emission
appears stronger in these areas. From Figure \ref{fig:optic_con}, it can be seen that
source 6, as mentioned above, is coincident with an area of optical emission at the centre
of the bridge. It can also be seen that an X-ray source (source 5) to the south of NGC 3396 has an optical counterpart. The
spectral model for this source was an assumed powerlaw due to the low
number of counts. Although the model used indicates that the source could
be an XRB, it is also possible that it could be a
supernova. From \citet{Huang_94} it was found that the last episode of
widespread star formation throughout the system took place
$\sim$4$\times$10$^{7}$yr ago, which is greater  than the timescale for
massive stars to evolve into supernovae ($\sim$1$\times$10${^7}$yr). Two of
the characteristics used to identify sources as XRBs are the
displacement of a source from an optical counterpart and evidence of
variability, neither of these are found with source 5. However, the
absence of these characteristics does not imply that the source is a supernova,
as kickouts do not always occur, and the low number of counts may
account for the lack of variability observed. In fact, due to the poor
statistics for this source it is impossible to rule out either
scenario. Sources 5 and 6 are the only two X-ray sources that have optical
counterparts and are external to the central regions of the two
galaxies. 

Comparing all the multi-wavelength observations it is clear
that the centres of both galaxies are sites of vigorous star
formation. It is also apparent that the disks of the two galaxies have
recently collided, resulting in an active star forming region. It is likely
that the bridge observed in both the optical and radio data has been
formed by tidal interactions.

%
\subsubsection{The luminosity function}

\begin{figure}
    \includegraphics[width=\linewidth]{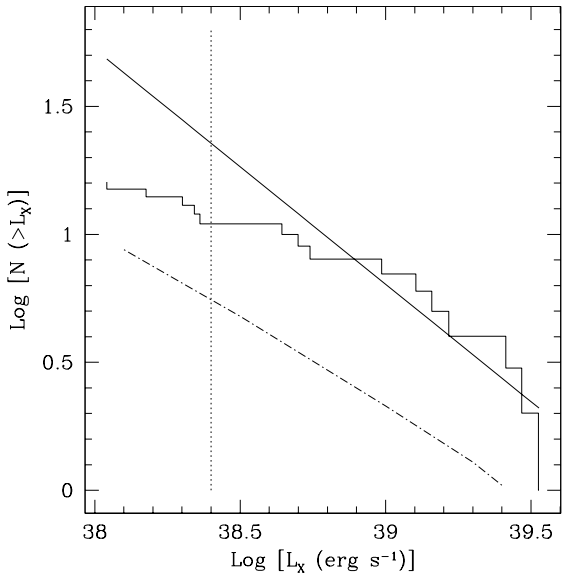}

    \caption{The Luminosity Function of the discrete sources of NGC
  3395/96. Plotting log$_{10} N$ against log$_{10}$ \LX, where $N$ is the cumulative number of sources above that luminosity, after correcting for incompleteness. The solid line indicates a power law fit to to the data giving a slope of -0.92 $_{-0.04}^{+0.76} $. The theoretical LF from \citet{Grimm_03} is also included (dot-dashed line); this gives a slope of -0.71. The dotted line indicates the lower cut-off luminosity, log \LX=38.4. } 
    \label{fig:Luminosity}

\end{figure}

Using the calculated intrinsic luminosity of each of the 16 sources, a
luminosity function (LF) was constructed and fitted with {\em
  Sherpa}, using the polynom1d
model.\footnote{http://cxc.harvard.edu/ciao/ahelp/polynom1d.html} This
power law fit of log $N$ against log \LX, where $N$ is the cumulative number of sources above that luminosity, gives a poorly constrained value of slope (-0.64 $_{-0.04}^{+0.44} $). After correcting for source incompleteness by excluding the lower luminosity sources from the fit, setting a conservative limiting value of log \LX=38.4 for the cut-off luminosity, a slope of (-0.92$_{-0.04}^{+0.76}$) was attained. This luminosity threshold has been based on our detection threshold of 1.4 $\times 10^{38}$ erg s$^{-1}$ for sources external to the body of the galaxies. This value rises in diffuse gas and to allow for this variation a conservative luminosity threshold of 2.5 $\times 10^{38}$ erg s$^{-1}$ has been adopted. The cut-off luminosity fit was again poorly constrained and from visual inspection it is apparent that the data requires a broken power law model. However, the statistics of the LF are unable to constrain this fit. \citet{Grimm_03} proposed the existence of a universal luminosity function of HMXRBs, with a normalisation proportional to the star formation rate (SFR) of a galaxy, given by
\begin{equation}
N(>L)=5.4\mathrm{SFR}(L_{38}^{-0.61}-210^{-0.61}),
\end{equation}
where $L_{38}$ = $L/10^{38}$ erg s$^{-1}$, SFR was calculated using the expression \citep{Rosa_02}
\begin{equation}
\mathrm{SFR}_\mathrm{FIR}=4.5 \times 10^{-44}L_\mathrm{FIR}(\mathrm{erg~s}^{-1}).
\end{equation}
From these expressions we find the \citet{Grimm_03} LF to have a slope of -0.71 for Arp 270. This theoretical LF is compared to the cumulative LF after correcting for source incompleteness in Figure \ref{fig:Luminosity}. The solid line shows the fit to the data and the dot-dashed line indicates the \citet{Grimm_03} LF. Although the LF slopes are in agreement, within errors, the normalisation is lower for the theoretical line, indicating that galaxy SFR may not be a good normalisation factor for LFs. In fact, in Figure \ref{fig:LFIR} we have shown that \NULX\ per galaxy is not directly proportional to \LFIR, a measure of SFR. 

\citet{Colbert_03} have found that LFs in merger and irregular galaxies have quite similar slopes to those in spiral galaxies, but that elliptical galaxies exhibit steeper slopes, finding a mean of -0.65$\pm $0.16 for the merger and irregular galaxies compared to -0.79$\pm $0.24 for spirals and -1.41$\pm $0.38 for ellipticals. In Table \ref{tab:compare} all 3 values are consistent with a slope of $\sim$-0.6, noting the large errors in the Arp 270 LF fit.

\subsection{Diffuse Gas}

As described in section \ref{sec:pobs} of this paper, Arp 270 has been
observed in X-rays before. The most recent of these observations are
detailed in RP98. From the \ROSAT\ data, the total luminosity of Arp
270 was found to be 3.1 $\times 10^{40}$erg s$^{-1}$ (0.1$-$2.0 keV), very similar to
the findings of this paper, which gives a total luminosity of
3.0$\times 10^{40}$erg s$^{-1}$ (0.3$-$6.0 keV). The difference in values can be
attributed to the different energy ranges that have been used
and the omission of the flux from a quasar in close proximity to Arp
270. This object was included in the \ROSAT\ data but excluded from
this study, as it lies $\sim$ 1\arcm\ outside the D$_{25}$ ellipses. It
was found in RP98 that 51\% of the total emission was made up of
diffuse gas, whereas the present analysis shows that only 32\% of the
total luminosity arises from diffuse gas. This result demonstrates 
the greater resolving power of \CHANDRA\ to resolve out point sources from diffuse emission.

\subsubsection{Temperature variation}

\begin{figure*}
  \includegraphics[width=14cm]{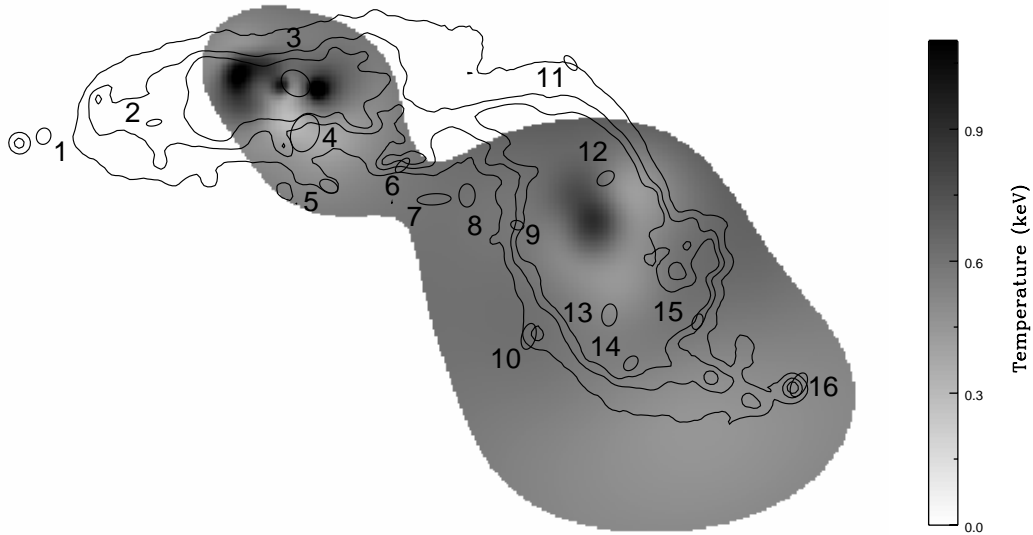}
  \hspace{0.5cm}

 \caption {A hardness map of Arp 270, the point sources having been removed, with optical contours and region
 files detected in the 0.3-8\,keV band. The dark features indicate
 hard X-ray emission.}
  \label{fig:hmap}
\end{figure*}

As mentioned above (\textsection \ref{sec_diffuse}), it
is not realistic to assume that the diffuse gas has a single
temperature. To investigate the spatial variation of temperature, a
hardness map has been created as a proxy for temperature (Figure
\ref{fig:hmap}). This map was created by taking a ratio of the medium
and soft bands of the adaptively smoothed diffuse gas, using the same set of smoothing scales, after background subtraction and point source removal. Features seen in NGC 3396 in the diffuse gas (Figure
\ref{fig:dif_gas}) and the variation map (Figure \ref{fig:vmap}) are also observed to have counterparts in the hardness map. These sources of emission were not formally identified in {\em wavdetect} as point sources, but the
evidence presented in this paper indicates that they are compact
objects, as they all show variability on a six month timescale.  The
hardest such feature in NGC 3396 has a temperature in excess of 1.1\,keV.
The diffuse gas across the rest of the galaxy pair emits at $\sim$0.5\,keV,
consistent with diffuse gas temperatures from other merging systems
(RP98), with hotter emission of $\sim$0.7\,keV in the regions of most active star formation.

\subsubsection{Physical properties of the diffuse gas}

From the spectral fits of the diffuse gas, and some assumptions about the
geometry of the emitting region, physical properties of the gas can be derived. Firstly the volume, $V$, of the emitting region of each galaxy has
been assumed to be an ellipsoid, with symmetry about the
major-axis. The fitted emission measure is equal to $\eta n_\mathrm{e}^{2}V$
and can be used to infer the mean particle number density $n_\mathrm{e}$, with
the filling factor, $\eta$, assumed to be 1. This factor represents the fraction of volume filled by the emitting gas. Although we have assumed this to be 1 in our calculations, there is evidence from hydrodynamical simulations to suggest this value could be $\le$2 per cent \citep{Strickland_00b}. Corrections must be applied to
the volume to account for the reduction in area due to the removal of
point sources. The mean electron density is then used to derive the
total gas mass $M_\mathrm{gas}$, which leads to the thermal energy $E_\mathrm{th}$,
and cooling time $t_\mathrm{cool}$, of the hot gas. These parameters can be
seen in Table \ref{tab:param}, alongside values given in \citet{Read_03} for the two galaxies that comprise The Mice. Although The Mice is likely at a later stage in the merger process than Arp 270, it is believed \citep{Read_03} that the diffuse emission from the galaxies does not evolve as rapidly as the point source population. It is therefore useful to compare the diffuse gas parameters of both galaxy systems.
\begin{table*}
\begin{center}
\caption{Comparing the physical parameters of Arp 270 and The Mice
derived from the diffuse gas. Here an ellipsoid model is assumed, $\eta$ is the filling factor.
\label{tab:param}}
\begin{tabular}{cc@{}c@{}ccccc@{}c@{}}    
 \noalign{\smallskip}
 \hline

Galaxy &Galaxy & $kT$ & Semi & Semi  & $n_\mathrm{e}$                & $M_{\mbox{\small gas}}$ & $E_{\mbox{\small th}}$ & $t_{\mbox{\small cool}}$  \\
System & & & Major-axis      &  Minor-axis        & (cm$^{-3}$           ) & ($M_{\odot}$)           & (erg)                  & (Myr)                     \\
       & & (keV)     & (kpc)           &  (kpc)            &($\times1/\sqrt{\eta}$) & ($\times\sqrt{\eta}$)   &  ($\times\sqrt{\eta}$) & ($\times\sqrt{\eta}$)     \\ 
\hline
Arp 270  & NCG 3395  & 0.52 & 3.3 & 3.1 & 0.051 & 10.20$\times10^{7}$ & 2.5$\times10^{56}$ & 1432 \\
Arp 270  & NGC 3396  & 0.49 & 2.9 & 1.3 & 0.122 & 4.0$\times10^{7}$  & 0.9$\times10^{56}$ & 571  \\
The Mice & NGC 4676A & 0.50 & 3.2 & 3.2 & 0.019 & 6.5$\times10^{7}$  & 1.8$\times10^{56}$ & 480  \\
The Mice & NGC 4676B & 0.46 & 2.6 & 2.6 & 0.025 & 4.3$\times10^{7}$  & 1.2$\times10^{56}$ & 380  \\
\hline

\end{tabular}

\end{center}
\end{table*}
 From this table it can be seen that the temperature of the diffuse gas in all four galaxies agrees well with the expected temperature of $\sim$ 0.5\,keV found for diffuse gas in interacting galaxies (RP98). By looking at the mean electron density, it can be seen
that NGC 3396 has a significantly higher value than the other three
galaxies. This is likely to be caused by the molecular and HI gas content of the galaxy and its high inclination from the line of sight, it could also be the result of emission from the suspected point sources at the centre of the galaxy. Also from Table \ref{tab:param}, it can be seen that the cooling times for both galaxies are much greater than the time since the last widespread starburst took place. Even if the low filling factor is included, the cooling times are still greater by a factor of 2, or more. 

\subsubsection{Origins of the diffuse X-ray emission}

From \citet{Mattila_01} a relationship between the supernova rate of starburst galaxies and \LFIR\ has been calculated:
\begin{equation}
\rSN=2.7\times10^{-12}\LFIR/\Lsol\ \mathrm{yr}^{-1},
\label{eqn:rsn}
\end{equation}
generally, \rSN\ is expressed in the SN unit (SNu), i.e. the number of supernovae per century per 10$^{10}$\Lsol\ of blue luminosity. However, in more active galaxies, it is more useful to normalise the observed number of supernovae to \LFIR,  which is known to be proportional to the galaxy star formation rate \citep{Mattila_01,Mannucci_03}. Using equation \ref{eqn:rsn}, \rSN\ for Arp 270 is calculated to be 0.036 SNyr$^{-1}$. Assuming that a massive star takes $\sim1\times 10^7$yr to evolve into a supernova and that the last episode of widespread star formation took place $4\times 10^7$yr ago \citep{Huang_94}, we calculate the number of SNe which have occurred within the galaxy pair to be $\sim1\times$10$^6$. Hence, assuming that each supernova will release $10^{51}$erg of energy, the total thermal energy available is $\sim1\times 10^{57}$ erg. This exceeds the value of 3.4$\times 10^{56}$ erg we have calculated from spectral modelling of the diffuse gas, although there will be additional energy sinks in the form of gravitational potential energy, wind kinetic energy and radiative losses. 

Another possible source of diffuse X-ray emission is unresolved XRBs. We can estimate the X-ray luminosity arising from this contribution by extrapolating the X-ray luminosity below our detection threshold limit of 1.4$ \times$10$^{38}$erg s$^{-1}$ using the universal luminosity function derived by \citet{Grimm_03}. This gives $L_\mathrm{Total}\approx$1.2$\times$10$^{39}$erg s$^{-1}$, for  10$^{37}$erg s$^{-1}\leq$\LX $\leq 1.4\times10^{38}$erg s$^{-1}$, thus showing that XRBs may be responsible for up to $\sim$12\% of this seemingly diffuse emission. To allow for this XRB contribution we have re-fitted the diffuse gas for both galaxies including an additional powerlaw contribution, with fitted values of \NH\ and a photon index fixed at $\Gamma$=1.5; a typical XRB spectrum \citep{Soria_03}. The normalisation for this component has been frozen to limit the XRB contribution to $\sim0.6\times10^{39}$erg s$^{-1}$ for each galaxy. This spectral model resulted in lower temperature fits for both NGC 3395 and NGC 3396 of 0.51$_{-0.15}^{+0.06}$ keV and 0.39$_{-0.04}^{+0.07}$ keV respectively, although these values are consistent, within errors, with the previously derived temperatures (Table \ref{tab:gas}). The intrinsic X-ray luminosities arising from the diffuse gas in the galaxies are now found to be 3.37$\pm 0.16 \times 10^{39}$erg s$^{-1}$ for NGC 3395 and 4.63$\pm 0.23 \times 10^{39}$erg s$^{-1}$ for NGC 3396. Deriving the physical parameters of Arp 270 with these new, fitted, values results in a reduction in $M_\mathrm{gas}$ of $\sim 2\%$ and $E_\mathrm{th}$ by $\sim4\%$ for NGC 3395. The reduction in NGC 3396 for $M_\mathrm{gas}$ is $\sim1\%$, but $E_\mathrm{th}$ is lower by $\sim25\%$, this is due to the lower fitted temperature derived from our spectral fit including an XRB contribution.

\subsubsection{Comparisons with other merging systems}

A comparison to The Mice highlights the absence of any gaseous outflows from Arp 270. In The Mice, diffuse outflows are seen coming from both galaxies. These features are small and are likely to be young starburst driven winds escaping from the nuclei of the galaxies. Arp 270 does not show these features, presumably since the gas has not had enough time to break out of the galaxy disks to form an outflow.

Evidence from previous studies (RP98) suggests that the galaxy pair NGC
3395/3396 is at an early stage of interaction. \citet{Clemens_99} have observed an HI tail extending from the south-east of the galaxy pair, it is likely that this feature was stripped from NGC 3395 during the system's first closest approach. This interpretation is strengthened by the lack of optical tidal tails in the system, as well as the absence of gaseous outflows. In more
advanced stages of galaxy evolution these features are good indicators
of strong galaxy interactions \citep{Toomre_72}.  

Although it is known that Arp 270 is an early merger candidate, just where in this sequence it lies has not been defined. From N-body simulations it has been shown that Arp 270 is within approximately 5$\times$10$^7$ years of its second perigalactic passage \citep{Clemens_99}. By comparing this to N-body simulations carried out on the other merger systems in our sample, a merger time sequence can be established. We can then compare the
properties derived from this {\em Chandra} observation to the merger
systems mentioned previously. Table \ref{tab:compare} shows a summary
of some key properties of these systems. Column (1) gives
the name of each system, (2) the time that has elapsed since the last perigalactic passage, (3) the overall X-ray luminosity
of the system, (4) the \LFIR/\LB\ ratio, (5) the percentage of
luminosity arising from the diffuse emission, and column (6) gives the slope of the
luminosity function.
\begin{table*}

\begin{center}
\caption{A comparison of the luminosities and galaxy properties of a
sample of interacting systems.}
\label{tab:compare}

\begin{tabular}{c@{}cc@{}cc@{}c}    \hline
Galaxy & Time Since First & \LX &  \LFIR/\LB & \% of & LF   \\
System & Perigalactic Passage$^\mathrm{a}$ & &  & & \\

       & Myr & $ \times 10 ^{40}$ erg s$^{-1} $ & &  Diffuse Emission & Slope \\
 
 \noalign{\smallskip}
 \hline

NGC 4485/90  & $\sim$40$^\mathrm{b}$    & 2.00 & 0.55  &  10 & -0.57 $        \pm $ 0.10  \\
Arp 270      & $\sim$50$^\mathrm{c}$    & 2.95 & 0.81  &  32 & -0.92 $ _{-0.04}^{+0.76} $ \\
The Mice     & $\sim$180$^\mathrm{d}$   & 7.03 & 1.13  &  25 &  -                         \\
The Antennae & $\sim$210$^\mathrm{d}$   & 8.40 & 0.94  &  45 & -0.53 $        \pm $ 0.07  \\
NGC 3256     & $\sim$500$^\mathrm{e}$   & 35   & 5.83  &  80 & -                          \\

\hline

\end{tabular}

\end{center}
Notes: $^\mathrm{a}$Based on N-body simulations.$^\mathrm{b}$\citet{Elmegreen_98},  $^\mathrm{c}$\citet{Clemens_99}, $^\mathrm{d}$\citet{Mihos_93}, $^\mathrm{e}$\citet{English_03}.
\end{table*}

By comparing the time-scales from N-body simulations in Table \ref{tab:compare}, it can be seen that Arp 270 lies second in this merger sample. Although, from the same N-body simulations, it was shown that an initial interaction between the two galaxies has taken place $\sim$5 $\times$ 10$^8$ years ago, this encounter was a weak fly-by interaction, unlike the strong NGC 3256 merger encounter which took place at the same time. We therefore characterise the timescale since the first strong interaction in Arp 270 as the time of $\sim$5 $\times$ 10$^7$ years since its second perigalactic passage.
With the star formation activity,  \LFIR/\LB, a problem we draw attention to is that, during recent star formation,
\LB\ is greatly enhanced and therefore not a good measure of the
galaxy mass. A better indicator would be red luminosity, but these
values are not readily available for this sample. Even so, it can be
seen in Table \ref{tab:compare} that, generally, \LFIR/\LB\ does
increase with evolution. Again, we see that The Mice appears to
be more active than The Antennae although we know, from N-body simulations, that The Antennae is
more evolved. Another useful indicator of evolutionary
stage is the percentage of luminosity arising from diffuse gas. \citet{Read_01} found that the more evolved the system (up to the ultraluminous, core merger stage), the higher the percentage. From Table \ref{tab:compare} it can be seen this
parameter gives support to Arp 270 being in a later stage of evolution
than NGC 4485/90, but earlier than The Antennae and NGC 3256.


\section{Conclusions}
\label{conclusions}

In this paper the \CHANDRA\ observations of the interacting galaxy
pair NGC 3395/3396 have been presented, along with previously
unpublished \ROSAT\ HRI observations. The two galaxies are thought to
be of comparable masses, and lie at a distance of 28 Mpc. The pair are at an
early stage of interaction, with the galaxy disks just starting to
collide. The most interesting points from this paper can be summarised
as follows:

\begin{itemize}

\item{A total of 16 point sources were detected, 15 of these were
within the D$_{25}$ ellipse, with a further source lying less than
10\arcs\ outside the optical confines of the galaxy and obviously associated with an optical feature of NGC 3395. This is 12 more
than previously detected with \ROSAT. These sources were spectrally fitted using
either an absorbed power law model or absorbed MEKAL model.  Spectral
fits suggest that the X-ray population is predominantly made up of
XRBs, which is further supported by the observed variability of 7 of these sources. Given the time since the last starburst $\sim$4 $\times 10^7$
years ago \citep{Huang_94} these are mostly expected to be HMXRBs.}

\item{Of the detected sources, 7 exceed the Eddington limit for a
7\Msol\ star and are therefore classified as ULXs. The findings of \citet{Swartz_04} indicate that there is a correlation between the number of ULXs and \LFIR. Further evidence of this is given in this paper: in Figure \ref{fig:LFIR}  the slope of the relationship is found to be \NULX$\propto\LFIR^{0.18}$. We also find that the more active systems host more luminous ULXs. This is indicated by the steeper slope exhibited in our \LULX\ against \LFIR\ fit where the slope is \LULX$\propto \LFIR^{0.54}$. These results, alongside the findings from \citet{Colbert_03} of the relation between point source X-ray luminosity and $L_\mathrm{K}$ with $L_\mathrm{FIR+UV}$, indicate that ULX populations are made up of a mixture of old and young sources.}

\item{X-ray features are found to coincide with the star
forming knots found in the FUV data \citep{Hancock_03}. The formally detected source, source 3, resides near the centre of NGC 3396 and has an offset of 1.5\arcsec, corresponding to 210 pc. This offset could indicate that the source is a runaway binary, a HMXRB which has been ejected during the formation of the neutron star. }

\item{The luminosity function of the discrete sources, after correcting for source incompleteness, gives a slope of -0.92$^{+0.76}_{-0.04}$, which, within errors, is comparable to the value given by \citet{Colbert_03} of -0.65$\pm0.16$. This fit has been compared to a theoretical LF from \citet{Grimm_03} and, although the LFs have similar slopes, the theoretical line appears to suppress the number of discrete sources, indicating that SFR may not be a good normalisation factor for LFs.}

\item{The morphology of the diffuse gas in NGC 3395 is less
compact than NGC 3396, which has some sub-structure running
along its major-axis (i.e. in the east-west direction), and is highly inclined to the line of sight. In more evolved systems such as The Mice, hot gaseous outflows have been observed. We detect no such features in Arp 270, suggesting that the gas has not yet become compressed enough to create starburst-driven galactic winds. A bridge between the two galaxies has been detected in the optical and radio data. At the centre of this bridge, between the two galaxies, an X-ray source (source 6) has been detected. It is likely that this source is being triggered by the recent collision of the two galaxy disks.}

\item{The total (0.3$-$6.0 keV) X-ray luminosity of Arp 270 is 3.0$\times 10^{40}$erg
s$^{-1}$, with 32\% of the emission arising from the diffuse gas. This
is a smaller percentage of diffuse gas than more evolved merging
systems. Less diffuse gas is expected in earlier systems as this
emission is due to galactic winds and fountains which arise from a large
collection of supernovae. These supernovae can take up to a few $\times
10^7$ years to evolve and merge to form energetic superbubbles. From numerical simulations it is thought that Arp 270 is an early stage merger system, more evolved than NGC 4485/90 but not as evolved as The Mice, The Antennae or NGC 3256.}
\end{itemize}

From this study we have been able to establish that Arp 270 is an example of a young merger candidate. By comparing the diagnostics of this system to other interacting galaxies we have found that the system is consistent with trends identified in RP98. The \LX:\LFIR\ value measured in this paper is in agreement with the \LX:\LFIR\ relation shown in RP98. We find that 32\% of the X-ray emission from this system arises from the diffuse gas. This small percentage is expected in such a young merger system as the gas has not yet been compressed into the nuclei of the galaxies. This is further compounded by the absence of any hot gaseous outflows. These results also show that Arp 270 has a significant population of ULX sources, with
almost 50\% of the discrete X-ray sources detected exceeding 1$\times 10^{39}$ erg
s$^{-1}$. From spectral fitting it is likely that the ULXs are XRBs, an indication which is given further credence by
the observed variability of four of these sources.


\section{Acknowledgements}

We thank the \CHANDRA\ X-ray Center (CXC) Data Systems and Science
Data Systems teams for developing the software used for the reduction (SDP) and analysis
(CIAO). We would also like to thank the anonymous referee for helpful comments which improved this paper, and Mark Hancock for providing the FUV FITS files
for Arp 270. NJB acknowledges the support of a PPARC studentship.


\label{lastpage}

\bibliographystyle{mn2e}
\bibliography{njb}

\end{document}